A

# Design of a High-Performance GEMM-like Tensor-Tensor Multiplication

Paul Springer, AICES, RWTH Aachen
Paolo Bientinesi, AICES, RWTH Aachen

We present "GEMM-like Tensor-Tensor multiplication" (GETT), a novel approach for dense tensor contractions that mirrors the design of a high-performance general matrix-matrix multiplication (GEMM). The critical insight behind GETT is the identification of three index sets, involved in the tensor contraction, which enable us to systematically reduce an arbitrary tensor contraction to loops around a highly tuned "macro-kernel". This macro-kernel operates on suitably prepared ("packed") sub-tensors that reside in a specified level of the cache hierarchy. In contrast to previous approaches to tensor contractions, GETT exhibits desirable features such as unit-stride memory accesses, cache-awareness, as well as full vectorization, without requiring auxiliary memory. We integrate GETT alongside the so called Transpose-Transpose-GEMM-Transpose and Loops-over-GEMM approaches into an open source "Tensor Contraction Code Generator" (TCCG). The performance results for a wide range of tensor contractions suggest that GETT has the potential of becoming the method of choice: While GETT exhibits excellent performance across the board, its effectiveness for bandwidth-bound tensor contractions is especially impressive, outperforming existing approaches by up to $12.4\times$. More precisely, GETT achieves speedups of up to $1.41\times$ over an equivalent-sized GEMM for bandwidth-bound tensor contractions while attaining up to 91.3% of peak floating-point performance for compute-bound tensor contractions.

## 1. INTRODUCTION

Dense tensor contractions (TC) are a common and critical component of scientific computations, encountered in fields as diverse as machine learning [Abadi et al. 2015; Vasilache et al. 2014], spectral element methods [Tufo and Fischer 1999], quantum chemistry calculations [Harrison et al. 2016; Bartlett and Musiał 2007], multidimensional Fourier transforms [Frigo and Johnson 2005; Pekurovsky 2012] and climate simulations [Drake et al. 1995]. Despite the close connection between matrix-matrix products (GEMM) and TCs, the performance of the latter is in general vastly inferior to that of an optimized GEMM. To close such a performance gap, we propose a novel approach, **GEMM**-like **T**ensor-**T**ensor multiplication (GETT),[1] that captures the essential features of a high-performance GEMM. In line with a highly optimized GEMM, and in contrast to previous approaches to TCs, our high-performance implementation of the GETT approach is fully vectorized, exploits the CPU's cache hierarchy, avoids explicit preprocessing steps, and operates on arbitrary dimensional sub-tensors while preserving stride-one memory accesses.

From a computational perspective, tensors can be interpreted as higher dimensional matrices or simply as multidimensional arrays; likewise, tensor contractions are a generalization of the matrix-matrix multiplication to higher dimensions. Given this connection, the disparity in performance between optimized GEMMs and TCs is striking. While GEMM implementations typically attain a significant fraction of the system's peak floating-point performance, this is often not the case for TCs (see Sec. 7 for a detailed discussion).

The lower efficiency of TCs can be mostly attributed to the escalated complexity due to the increased dimensionality of tensors (e.g., 6D [Apra et al. 2014], or 8D [Kucharski and Bartlett 1992; Baumgartner et al. 2005]), which often results in suboptimal memory accesses; as a consequence, the performance of many tensor contractions can be limited by the system's memory-bandwidth (i.e., bandwidth-bound) as opposed to its floating-point units (i.e., compute-bound). Hence, developing a reliable and systematic way to exploit the system's caches is critical in order to increase the performance of TCs and push them from the bandwidth-bound regime to the compute-bound regime [Williams et al. 2009].

Previous research on tensor contractions can be mainly classified around three approaches: 1) (vectorized) nested-loop code, 2) **T**ranspose-**T**ranspose-**G**EMM-**T**ranspose (TTGT), and more recently, 3) **L**oops-**o**ver-**G**EMMs (LoG); each approach is discussed below.

*Nested loops.* Implementations based on nested loops [Apra et al. 2014] improve the performance over the direct translation of the mathematical definition by applying loop transformations (e.g., loop-reordering, loop-fusion). These methods typically suffer from strided memory access, thus causing suboptimal memory access patterns and low utilization of the memory subsystem. Stock et al. [Stock et al. 2011; Stock et al. 2012a] present a sophisticated vectorizations scheme for small tensor contractions which fit into the caches [Harrison et al. 2004; Harrison et al. 2016]. While the downsides of the purely nested-loop implementations can be mitigated by vectorization, this strategy is not sufficient for larger tensor contractions which do not fit into the caches and, therefore, require some sort of memory optimization to improve the cache utilization. Ma et al. [Ma et al. 2011] presented a code generator which generates high-performance CUDA code—based on the loop-based approach—for the tensor contractions arising in the so called regularized CCSD(T) method [Kowalski and Valiev 2009]. Another loop-based code generator for GPUs was later also developed by Nelson et al. [Nelson et al. 2015].

*Transpose-Transpose-GEMM-Transpose.* The key idea behind TTGT [Hirata 2003] is to exploit the highly efficient GEMM implementations offered by tuned BLAS libraries (e.g., ATLAS [Whaley and Petitet 2005], OpenBLAS [Wang et al. 2013], BLIS [Van Zee and van de Geijn 2015], MKL [Intel Cooperation 2016]). This method requires a preparation step which "flattens" (or "unfolds") the arbitrary dimensional tensors into matrices—via explicit tensor transpositions—so that the contraction can be cast as a single GEMM; finally, the resulting matrix is folded into the desired tensor layout— again incurring an additional overhead. While this method has the potential to yield high-performance for compute-bound TCs [DePrince III and Hammond 2011], it suffers from two inherent disadvantages: first, the transposed tensors require additional memory,[2] and second, the transposition process accounts for pure overhead. In many cases, this overhead dominates the runtime and renders the approach infeasible in the bandwidth-bound regime. Prominent adopters of TTGT are the Cyclops Tensor Framework (CTF) [Solomonik et al. 2013], Tensor Toolbox [Bader et al. 2012; Kolda and Bader 2009], Tensorlab [Vervliet et al. 2016] and libtensor [Epifanovsky et al. 2013]. The Tensor Contraction Engine (TCE) [Hirata 2003; Baumgartner et al. 2005] is a code generator adopted by the computational chemistry software package NWChem [Bylaska et al. 2016] to carry out coupled cluster calculations [Crawford and Schaefer 2000]; its generated code is a mixture of LoG and TTGT in the sense that it uses LoG at the distributed-memory level and TTGT for the shared-memory tensor contractions. In Sec. 4, we discuss different implementation candidates of TTGT and outline a performance metric by which the most promising candidate is selected. In contrast to other existing TTGT implementations, we rely on the Tensor Transpose Compiler (TTC)[3] to generate efficient tensor transpositions such that the pre- and post-processing overhead becomes less noticeable.

---

[1] We use the terms 'tensor-tensor multiplication' and 'tensor contraction' interchangeably.

*Loops-over-GEMMs.* Recent work on LoG [Di Napoli et al. 2014; Li et al. 2015] suggested to slice tensors into a sequence of 2D sub-tensors (matrices), and contract them via GEMMs. When the sizes of the sub-tensors involved are large, LoG is especially effective. However, depending on the TC, the 2D slices might be small and/or incur strided memory accesses, resulting in poor performance (this behaviour is illustrated in Sec. 7). Shi et al. [Shi et al. 2016]—independently of us—developed a CPU and GPU implementation following the LoG approach for *single-mode* contractions. They introduce the `stridedBatchedGemm` implementation to cuBLAS that is ideally suited for batches of small GEMMs with a constant stride between the matrices involved in all matrix-matrix multiplications. Moreover, they give guidelines for selecting a promising LoG candidate.

*GEMM-like Tensor-Tensor multiplication.* In this publication, we introduce GETT, a method that aims to capture the benefits of the aforementioned approaches, while avoiding their drawbacks. GETT is inspired by the work of Chetlur et al. [Chetlur et al. 2014] on convolutions in the context of machine learning. Convolutions, similarly to TCs, can be cast in terms of matrix-matrix multiplications if the operands are flattened into matrices—much like the TTGT approach for tensor contractions. However, in contrast to TTGT, Chetlur et al. avoid the costly preparation step before and after calling a GEMM by implicitly reorganizing the data while loading it into the caches. Similarly, the GETT approach reduces arbitrary tensor contractions to nested loops around a highly-tuned "macro-kernel", for which the prepared operands—multi-dimensional sub-tensors—reside in a specified level of the cache hierarchy.

GETT's design is motivated by previous research on high-performance matrix-matrix multiplications, where a large GEMM is reduced to a series of calls to a specialized (smaller) macro-kernel [Gunnels et al. 2001; Goto and Geijn 2008; Van Zee and van de Geijn 2015]; conceptually, GETT's macro-kernel is similar to that used in these high-performance GEMMs. Furthermore, GETT is akin to the TTGT approach with the critical difference that the overhead associated with the preparation of the tensors before and after calling GEMM is avoided; instead, GETT suitably prepares ("packs") sub-blocks (not necessarily two-dimensional) of the tensors into the CPU's caches as they are needed by the specialized macro-kernel, thereby reducing the data traffic to the slower main memory. Hence, one can alternatively think of GETT as a "fused TTGT" where the transpositions are "fused" into the GEMM such that GETT does not require additional memory, and does not suffer from the overhead due to the explicit transpositions prior to and after the contraction.

---

[2]The memory requirement is equal to the size of the transposed tensors. Thus, the memory footprint might increase by as much as 2×.
[3]The source code is available at www.github.com/HPAC/ttc.

Recently, Devin Matthews independently developed TBLIS [Matthews 2016], a C++ library for tensor contractions that also closely follows the design of a high-performance matrix-matrix multiplication—just like GETT. Both of our approaches share the same key insight, namely: The tensor transpositions need to be avoided and should be moved into a GEMM-like kernel. Aside from this similarity GETT and TBLIS set different emphases: First and foremost, GETT's packing routines are based on tensor transpositions, thus exploiting the spatial locality inherent to these operations. Second, GETT uses an auto-fine-tuning framework—guided by a performance model—to select among several implementations. TBLIS, on the other hand, supports all forms of tensor contractions, including those which cannot be mapped to a GEMM-like kernel [Di Napoli et al. 2014] and offers a library solution that does not require an extra compilation step.

The challenges behind a GEMM-like tensor-tensor multiplication are manifold: one has to identify arbitrary dimensional sub-tensors of appropriate size, and develop a systematic way to pack them into 2D or 3D contiguous tensors to increase spatial locality, so that they reside in a specified level of the cache hierarchy; furthermore, we address these challenges while ensuring that the stride-one index of these sub-tensors is preserved, such that the packing can be as efficient as possible. One of the most critical insights behind GETT is that the arbitrary dimensional sub-tensors (that are passed to the specialized macro-kernel) can be logically interpreted as higher (or lower) dimensional tensors so that they can be prepared (packed-and-transposed) via tensor transpositions. In light of this insight, our GETT implementation relies on the Tensor Transpose Compiler, which is guaranteed to use unit-stride memory accesses, irrespective of the actual transposition. Because of this, GETT avoids non-unit-stride memory accesses, regardless of the actual TC considered.[4]

In sharp contrast to previous approaches, GETT preserves the arithmetic intensity[5] for any given tensor contraction compared to an equally-sized GEMM; this property is critical for high performance. GETT's advantages—its preferable memory access pattern, its ability to pack data for the various levels of the cache hierarchy and its highly-tuned, vectorized macro-kernel—translate to an excellent performance signature, especially for bandwidth-bound TCs.

*Tensor Contraction Code Generator.* Our GETT, LoG and TTGT implementations are combined into a unified tool, the **T**ensor **C**ontraction **C**ode **G**enerator (TCCG). The speedups of single and double precision TCCG-generated code over other existing TTGT-based implementations for a range of tensor contractions (the full benchmark is described in Sec. 7.1) are illustrated in Fig. 1. In all cases, TCCG is at least as fast as the other approaches, attaining speedups between $1.0\times$ and $12.4\times$ in single precision (see Fig. 1a), and between $1.0\times$ and $3.7\times$ in double precision (see Fig. 1b).[6] As it will become apparent in Sec. 7, the low speedups, e.g. test cases 21–24, correspond to contractions for which the efficiency is very close to peak floating-point performance.

---

[4]Unless the size of the stride-one index of any tensor becomes one.
[5]The ratio of floating-point operations to data-related operations, see [Williams et al. 2009] for details.
[6]The speedups for single precision are higher than those of double precision because some reference TTGT-based implementations do not support single precision, see Section 7.

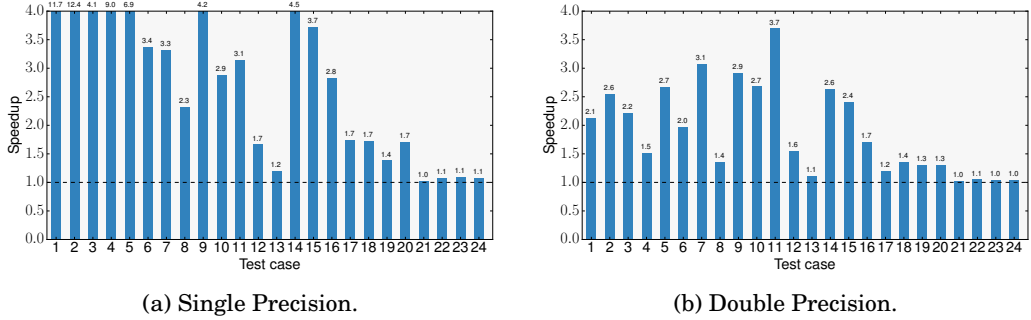

(a) Single Precision.  (b) Double Precision.

Fig. 1: Speedup of TCCG over the best reference version across a wide range of tensor contractions. Host: Intel *Xeon E5-2680 v3* running with one thread.

*Organization.* The remainder of this paper is structured as follows. Sec. 2 provides the necessary background information for this paper to be self-contained. Sec. 3 presents the core contribution and describes the GETT approach in detail. Secs. 4 and 5 cover the Transpose-Transpose-GEMM-Transpose and Loops-over-GEMM approaches, respectively. A unified code generator (the Tensor Contraction Code Generator) is introduced in Sec. 6, and an extensive performance evaluation is given in Sec. 7. Finally, Sec. 8 summarizes our work and outlines possible future directions.

## 2. BACKGROUND

To make this document self-contained, we provide here an introduction to general high-performance matrix-matrix multiplications, tensor contractions, and tensor transpositions.

### 2.1. Matrix-Matrix Multiplication

Let $A \in \mathbb{R}^{M \times K}$, $B \in \mathbb{R}^{K \times N}$, and $C \in \mathbb{R}^{M \times N}$ be input matrices; following the BLAS interface [Dongarra et al. 1990], a general matrix-matrix product is expressed as

$$\forall i \forall j \ \ C_{i,j} \leftarrow \sum_k \alpha \times A_{i,k} \times B_{k,j} + \beta \times C_{i,j}. \qquad (1)$$

Listing 1 contains the direct translation of Eq. 1 into code, in the form of three nested loops. Due to the poor exploitation of the caches (i.e., a lot of the data is redundantly fetched from main memory), such a naive implementation yields extremely poor performance. Several detailed discussions of high-quality GEMM implementations exist [Gunnels et al. 2001; Goto and Geijn 2008; Van Zee and van de Geijn 2015; Smith et al. 2014; Low et al. 2015]; here we only sketch the ideas underlying a high-performance GEMM, as preliminaries for the next sections.

```
// N-Loop
for j = 0 : N − 1
    // M-Loop
    for i = 0 : M − 1
        tmp = 0
        // K-Loop (contracted)
        for k = 0 : K − 1
            tmp += A_{i,k} B_{k,j}
        // update C
        C_{i,j} = α * tmp + β * C_{i,j}
```

Listing 1: Naive matrix-matrix multiply.

```
// N-Loop
for n = 0 : n_C : N − 1
    // K-Loop (contracted)
    for k = 0 : k_C : K − 1
        B̂ = identify_submatrix(B, n, k)
        // pack B̂ into B̃
        B̃ = packB(B̂) // B̃ ∈ ℝ^{k_C × n_C}
        // M-Loop
        for m = 0 : m_C : M − 1
            Â = identify_submatrix(A, m, k)
            // pack Â into Ã
            Ã = packA(Â) // Ã ∈ ℝ^{m_C × k_C}
            Ĉ = identify_submatrix(C, m, n)
            // matrix-matrix product: Ã × B̃
            macroKernel(Ã, B̃, Ĉ, α, β)
```

Listing 2: High-performance GEMM.

A necessary ingredient, underlying any high-performance GEMM (see Listing 2), is to organize the computation into blocks (as opposed to scalars), and to pack such blocks (i.e., sub-matrices of $A$, $B$ and $C$) into contiguous arrays that fit into specific cache levels. This technique improves locality, and thereby reduces both cache misses and translation lookaside buffer (TLB) misses.

In Listing 2, the blocks are denoted by $\widehat{A} \in \mathbb{R}^{m_C \times k_C}$, $\widehat{B} \in \mathbb{R}^{k_C \times n_C}$, and $\widehat{C} \in \mathbb{R}^{m_C \times n_C}$, while their packed counterparts are the auxiliary arrays $\widetilde{A}$, and $\widetilde{B}$;[7] the parameters $m_C, n_C$ and $k_C$ are chosen according to the sizes of the caches for a given CPU. Each block, once loaded in cache, is reused multiple times before being finally evicted; this reduces the need to fetch data redundantly from the slow main memory.

Once the sub-matrices $\widetilde{A}$ and $\widetilde{B}$ are prepared (packed), they are multiplied via a macro-kernel, which essentially is a highly-tuned GEMM, customized for operands that are known to reside in cache. Such a macro-kernel is the fundamental building block of earlier GEMM implementations [Gunnels et al. 2001; Goto and Geijn 2008]; more recent work of Van Zee et al. [Van Zee and van de Geijn 2015] proposes to break this building block down into even smaller micro-kernels, implemented in assembly, for maximum control of the CPU's resources.

Gunnels et al. [Gunnels et al. 2001] identified three variants to break down a matrix-matrix multiplication into a series of 1) matrix-panel, 2) panel-matrix, or 3) panel-panel multiplications (see Table I). Each of these variants corresponds to a different ordering of the loops in Listing 2.

### 2.2. Tensor Contractions

Let $\mathcal{A} \in \mathbb{R}^{S_1^\mathcal{A} \times S_2^\mathcal{A} \times \ldots S_{d_\mathcal{A}}^\mathcal{A}}$, $\mathcal{B} \in \mathbb{R}^{S_1^\mathcal{B} \times S_2^\mathcal{B} \times \ldots S_{d_\mathcal{B}}^\mathcal{B}}$, and $\mathcal{C} \in \mathbb{R}^{S_1^\mathcal{C} \times S_2^\mathcal{C} \times \ldots S_{d_\mathcal{C}}^\mathcal{C}}$ be $d_\mathcal{A}$-, $d_\mathcal{B}$- and $d_\mathcal{C}$-dimensional tensors, respectively. Extending the "contracted tensor product" of Bader et al. [Bader and Kolda 2006], we represent a tensor contraction as

$$\mathcal{C}_{\Pi^\mathcal{C}(I_m \cup I_n)} \leftarrow \sum_{k_1} \ldots \sum_{k_\xi} \alpha \times \mathcal{A}_{\Pi^\mathcal{A}(I_m \cup I_k)} \times \mathcal{B}_{\Pi^\mathcal{B}(I_k \cup I_n)} + \beta \times \mathcal{C}_{\Pi^\mathcal{C}(I_m \cup I_n)}, \qquad (2)$$

---

[7]$C$ is not packed.

|  | Condition | Shape |
|---|---|---|
| Matrix-panel multiplication | $n$ is small | $C$ = $A$ $B$ + $C$ |
| Panel-matrix multiplication | $m$ is small | $C$ = $A$ $B$ + $C$ |
| Panel-panel multiplication | $k$ is small | $C$ = $A$ $B$ + $C$ |

Table I: Shapes of a matrix-matrix multiplication; taken from [Gunnels et al. 2001].

where $\Pi^{\mathcal{A}}$, $\Pi^{\mathcal{B}}$ and $\Pi^{\mathcal{C}}$ are permutations[8] of the symbolic index sets $I_m := \{m_1, m_2, \ldots, m_\gamma\}$, $I_n := \{n_1, n_2, \ldots, n_\zeta\}$, and $I_k := \{k_1, k_2, \ldots, k_\xi\}$. These index sets respectively denote the free indices of $\mathcal{A}$ and $\mathcal{B}$ (i.e., those indices which appear in $\mathcal{C}$ and either $\mathcal{A}$ or $\mathcal{B}$), as well as the contracted indices (i.e., those indices which appear in both $\mathcal{A}$ and $\mathcal{B}$, but not in $\mathcal{C}$).[9] Notice that the following relations hold: $d_{\mathcal{A}} = \gamma + \xi$, $d_{\mathcal{B}} = \zeta + \xi$, and $d_{\mathcal{C}} = \gamma + \zeta$.

One critical observation is that by adopting the index sets $I_m$, $I_n$, and $I_k$, an arbitrary contraction can be represented in a GEMM-like fashion. Furthermore, to simplify the notation, in this manuscript we adopt the "Einstein Notation"[10] where the sums over the contracted indices are implicit. Eq. 2 becomes:

$$\mathcal{C}_{\Pi^{\mathcal{C}}(I_m \cup I_n)} \leftarrow \alpha \times \mathcal{A}_{\Pi^{\mathcal{A}}(I_m \cup I_k)} \times \mathcal{B}_{\Pi^{\mathcal{B}}(I_n \cup I_k)} + \beta \times \mathcal{C}_{\Pi^{\mathcal{C}}(I_m \cup I_n)}. \qquad (3)$$

***Example.*** Using this formalism, a general matrix-matrix multiplication can be expressed as follows: $I_m = \{m_1\}$, $I_n = \{n_1\}$, $I_k = \{k_1\}$ with $\Pi^{\mathcal{A}}(I_m \cup I_k) = (m_1, k_1)$, $\Pi^{\mathcal{B}}(I_k \cup I_n) = (k_1, n_1)$ and $\Pi^{\mathcal{C}}(I_m \cup I_n) = (m_1, n_1)$, yielding $\mathcal{C}_{m_1, n_1} = \mathcal{A}_{m_1, k_1} \mathcal{B}_{k_1, n_1}$.

In the following, we assume that $I_m$, $I_n$ and $I_k$ are not empty to express tensor contractions in terms of matrix-matrix multiplications; by contrast, when this assumption is violated, the contraction might be mapped onto lower level BLAS kernels (e.g., GEMV, DOT) [Di Napoli et al. 2014].

Moreover, we adhere to the Fortran memory layout: The indices of a tensor $\mathcal{T}_{i_1, i_2, \ldots, i_N}$ are stored from left to right (i.e., $i_1$ is the stride-one index). Similarly to the size $S_i \in \mathbb{N}$ of any index $i$, we use the same notation to express the size $S_I \in \mathbb{N}$ of an index set $I$ (i.e., $S_I := \Pi_{i \in I} S_i$).

### 2.3. Tensor Transpositions

Let $\mathcal{A} \in \mathbb{R}^{S_1^{\mathcal{A}} \times S_2^{\mathcal{A}} \times \ldots S_d^{\mathcal{A}}}$ be a $d$-dimensional tensor, $\alpha, \beta \in \mathbb{R}$, $\Pi^\star(i_1, i_2, \ldots, i_d)$ an arbitrary permutation, and $\widetilde{\mathcal{A}} \in \mathbb{R}^{\Pi^\star(S_1^{\mathcal{A}}, S_2^{\mathcal{A}}, \ldots, S_d^{\mathcal{A}})}$. A general tensor transposition, allowing both input and output scaling, is expressed as

$$\widetilde{\mathcal{A}}_{\Pi^\star(i_1, i_2, \ldots, i_d)} \leftarrow \alpha \times \mathcal{A}_{i_1, i_2, \ldots, i_d} + \beta \times \widetilde{\mathcal{A}}_{\Pi^\star(i_1, i_2, \ldots, i_d)}. \qquad (4)$$

---

[8]For the sake of readability, instead of using tuples to represent indices, we loosely use the terminology and the notation for sets ($\subset$, $\cup$ and $\cap$). Hence a permutation of an index set simply indicates that the indices can appear in any order.
[9]We do not consider the case in which one index appears in all tensors $\mathcal{A}$, $\mathcal{B}$ and $\mathcal{C}$.
[10]For the sake or readability, we do not differentiate covariant and contravariant indices and represent all indices as subscripts.

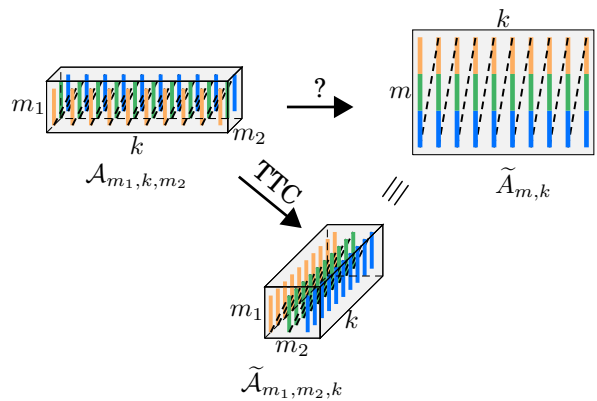

Fig. 2: Tensor transposition; visualization of flattening process. Lines illustrate the memory layout; solid lines represent contiguous chunks of memory.

It is no coincidence that we use the same notation ($\widetilde{\mathcal{A}}$) to indicate both the transposition and the packing of an operand ($\mathcal{A}$). Tensor transpositions can be used to flatten a tensor of arbitrary dimension down to matrix form, and thus are a critical component of the TTGT approach. Fig. 2 illustrates the case where the input tensor $\mathcal{A}_{m_1,k,m_2}$ (Fig. 2 top left) is flattened into the matrix $\widetilde{A}_{m,k} \equiv \widetilde{A}_{(m_1,m_2),k}$ (Fig. 2 top right) so that its size does not change (i.e., $S_m = S_{m_1} S_{m_2}$). This process can be carried out by a tensor transposition by reinterpreting the matrix $\widetilde{A}_{(m_1,m_2),k}$ as a 3D tensor $\widetilde{\mathcal{A}}_{m_1,m_2,k}$ (Fig. 2 bottom).

In a previous work, the authors introduced TTC, a compiler that generates explicitly vectorized and parallelized C++ code for any given tensor transposition [Springer et al. 2016a; Springer et al. 2016b]. By accepting a stride for each index, TTC can operate on sub-tensors; this feature makes TTC an ideal building block for GETT. As discussed later, TTC is used for the generation of high-performance packing routines for the sub-tensors of $\mathcal{A}$, $\mathcal{B}$ and $\mathcal{C}$.

## 3. GEMM-LIKE TENSOR-TENSOR MULTIPLICATION (GETT)

The key idea behind GETT is to reduce an arbitrary tensor contraction to a sequence of small matrix-matrix multiplications for which the operands (i.e. sub-tensors) fit into the caches. This approach is akin to the techniques used to implement a large GEMM in terms of smaller matrix-matrix multiplications which are computed by a 'macro-kernel' (compare Sec. 2.1). The observation is that the same way GEMM (Eq. 1) translates to the triple loop in Listing 1, an arbitrary tensor contraction (Eq. 3) can be computed by multiple nested loops, as shown in Listing 3—with the difference that GETT may require multiple $M$-, $N$- and $K$-loops. Notice that the update of the auxiliary variable `tmp` (Line 14) potentially requires scattered memory accesses to both $\mathcal{A}$ and $\mathcal{B}$.

tensor contractions expressed in terms of the index sets $I_m$, $I_n$ and $I_k$ (see Sec. 2.2) resembles the mathematical representation of a matrix-matrix multiplication (compare Eq. 2 and Eq. 1); these sets also enable one to express any tensor contraction in a similar way to a high-performance GEMM (compare Listings 2 and 4). Thus, GETT reduces an arbitrary dimensional tensor contraction to three loops around a macro-kernel, where the loop indices $m$, $n$ and $k$ respectively affect the free indices of $\mathcal{A}$ and $\mathcal{B}$

```
1   // N-Loops
2   for n_1 = 1 : S_{n_1}
3       // ... remaining N-loops omitted ...
4       for n_ζ = 1 : S_{n_ζ}
5           // M-Loops
6           for m_1 = 1 : S_{m_1}
7               // ... remaining M-loops omitted ...
8               for m_γ = 1 : S_{m_γ}
9                   tmp = 0
10                  // K-Loops (contracted)
11                  for k_1 = 1 : S_{k_1}
12                      // ... remaining K-loops omitted ...
13                      for k_ξ = 1 : S_{k_ξ}
14                          tmp += A_{Π_A(m_1,...,m_γ,k_1,...,k_ξ)} * B_{Π_B(k_1,...,k_ξ,n_1,...,n_ζ)}
15                  // update C
16                  C_{Π_C(m_1,...,m_γ,n_1,...,n_ζ)} = α * tmp + β * C_{Π_C(m_1,...,m_γ,n_1,...,n_ζ)}
```

Listing 3: Naive tensor-tensor multiplication.

```
1   // N-Loop
2   for n = 1 : n_C : S_{I_n}                                    ①
3       // K-Loop (contracted)
4       for k = 1 : k_C : S_{I_k}                                ②
5           B̂ = identify_subtensor(B, n, k)
6           // pack B̂ into B̃
7           B̃ = packB(B̂)                                        ③
8           // M-Loop
9           for m = 1 : m_C : S_{I_m}                            ④
10              Â = identify_subtensor(A, m, k)
11              // pack Â into Ã
12              Ã = packA(Â)                                     ⑤
13              Ĉ = identify_subtensor(C, m, n)
14              // matrix-matrix product: Ĉ ← αÃ × B̃ + βĈ
15              macroKernel(Ã, B̃, Ĉ, α, β)                      ⑥
```

Listing 4: High-performance GETT.

as well as the contracted indices.[11] To do so, one has to block the input operands along the $M$-, $N$- and $K$-dimensions, to create the packed, auxiliary arrays $\widetilde{\mathcal{A}} \in \mathbb{R}^{m_C \times k_C}$ and $\widetilde{\mathcal{B}} \in \mathbb{R}^{n_C \times k_C}$ which will be processed by the macro-kernel. While blocking improves the temporal locality, packing increases the spatial locality of the sub-tensors and—in contrast to the naive tensor-tensor contraction—avoids the scattered memory accesses to $\mathcal{A}, \mathcal{B}$ during the update of $\mathcal{C}$. Following this approach, the remarkable similarity between a high-performance GEMM (see Listing 2) and a high-performance GETT (see Listing 4) becomes evident.

In contrast to a naive implementation of a tensor-tensor multiplication where each index $i \in (I_m \cup I_n \cup I_k)$ appears as an individual loop (see Listing 3), a high-performance GETT (see Listing 4) replaces the multiple $M$-, $N$- and $K$- loops with just a single loop each. Hence, each loop counter (i.e., $m$, $n$ and $k$) potentially influences multiple indices of the tensors $\mathcal{A}, \mathcal{B}$ and $\mathcal{C}$. The exact mechanism is handled by the function identify_subtensor(), which depending on the current loop iteration, identifies the appropriate sub-tensors $\widehat{\mathcal{A}}, \widehat{\mathcal{B}}$ or $\widehat{\mathcal{C}}$ from either $\mathcal{A}, \mathcal{B}$ or $\mathcal{C}$; notice that this "identifi-cation process" happens entirely at compile time and does not cause any data to be moved. For now we assume that these sub-tensors are given, and that their sizes and dimensionality matches those of their packed counterparts $\widetilde{\mathcal{A}}, \widetilde{\mathcal{B}}$ and $\widetilde{\mathcal{C}}$; the function identify_subtensor() is discussed in detail in Sec. 3.2.

---

[11]To simplify the discussion, we only focus on one of the GEMM-variants listed in Table I, the matrix-panel multiplication.

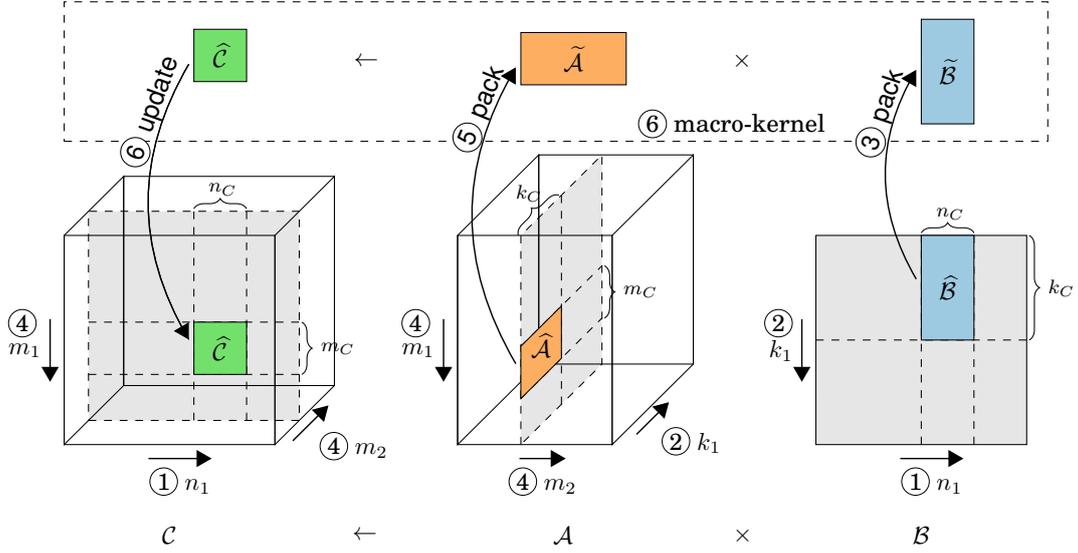

Fig. 3: GETT blocking overview for an exemplary tensor contraction $\mathcal{C}_{m_1,n_1,m_2} \leftarrow \mathcal{A}_{m_1,m_2,k_1} \times \mathcal{B}_{n_1,k_1}$. $\widehat{\mathcal{A}}, \widehat{\mathcal{B}}$ and $\widehat{\mathcal{C}}$ respectively correspond sub-tensors of $\mathcal{A}$, $\mathcal{B}$ and $\mathcal{C}$ which are not yet packed. The upper dashed box denotes the macro-kernel. The numbers correspond to those in Listing 4.

The program flow for an exemplary tensor contraction, $\mathcal{C}_{m_1,n_1,m_2} \leftarrow \mathcal{A}_{m_1,m_2,k_1} \mathcal{B}_{n_1,k_1}$, is illustrated in Fig. 3. Even though the sub-tensors are only two dimensional in this example, we stress that in the general case $\widehat{\mathcal{A}}, \widehat{\mathcal{B}}$ or $\widehat{\mathcal{C}}$ can be of arbitrary dimension; hence, these sub-tensors can be collected across multiple dimensions—not just two.

First, the ① n- and ② k-loops select a sub-tensor $\widehat{\mathcal{B}}$ of $\mathcal{B}$; this limits the size of the non-packed sub-tensor $\widehat{\mathcal{B}}$. $\widehat{\mathcal{B}}$ is then ③ packed into the contiguous auxiliary array $\widetilde{\mathcal{B}}$. The ④ m-loop follows; in the context of the example, this affects more than one index (i.e., $m_1$ and $m_2$), as opposed to the n- and k-loops, which only affect one index each (i.e., $n_1$ and $k_1$, respectively). Next, ⑤ the non-packed sub-tensor $\widehat{\mathcal{A}}$ of $\mathcal{A}$ is packed into the contiguous tensor $\widetilde{\mathcal{A}}$. Finally, the macro-kernel ⑥ is invoked to compute the matrix-matrix product of $\widetilde{\mathcal{A}}$ with $\widetilde{\mathcal{B}}$ to update the sub-tensor $\widehat{\mathcal{C}}$; notice that $\widehat{\mathcal{C}}$ is not packed and merely denotes a reference to the corresponding elements of $\mathcal{C}$.

### 3.1. Micro- and Macro-Kernel

Fig. 4 (bottom) illustrates the storage scheme of the auxiliary arrays $\widetilde{\mathcal{A}}$ and $\widetilde{\mathcal{B}}$.[12] The proper choice for the parameters $m_C$, $n_C$, $k_C$, $m_R$ and $n_R$ is discussed in Sec. 3.3; for now it suffices to know that these parameters are chosen such that $\widetilde{\mathcal{B}}$ remains in L3 cache, $\widetilde{\mathcal{A}}$ remains in L2 cache and that a $m_R \times k_C$ sub-matrix of $\widetilde{\mathcal{A}}$ alongside a $k_C \times n_R$ sub-matrix of $\widetilde{\mathcal{B}}$ simultaneously fit into the L1 cache. Thus, all packed sub-tensors will remain in some level of the cache hierarchy and will not require additional loads from main memory.

---

[12]This storage scheme is akin to the memory layout proposed by Van Zee et al. [Van Zee and van de Geijn 2015].

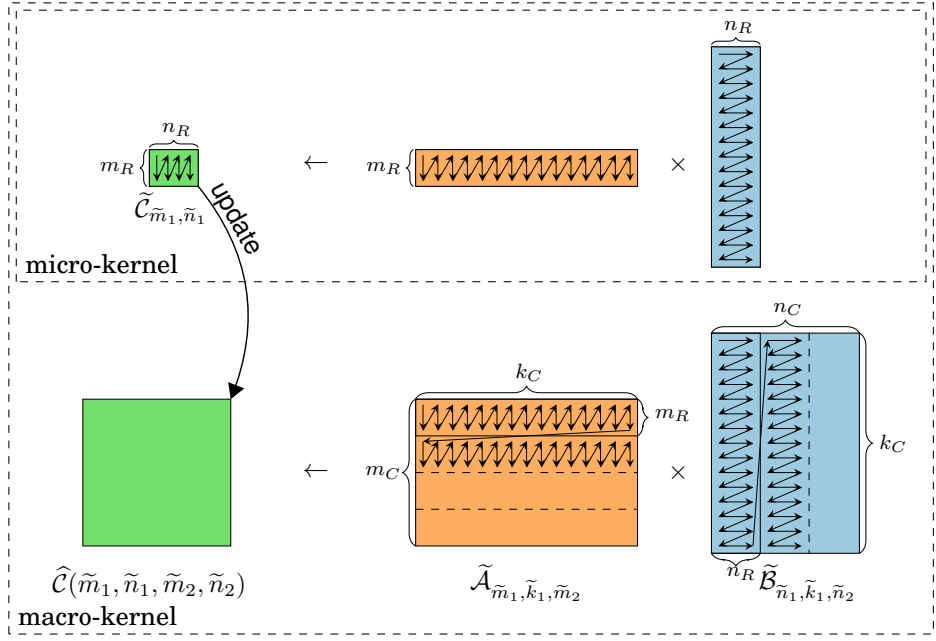

Fig. 4: Layout of the packed tiles $\widetilde{\mathcal{A}}$, $\widetilde{\mathcal{B}}$ and $\widetilde{\mathcal{C}}$. Arrows denote the storage scheme.

The macro-kernel computes the matrix-matrix multiplication of $\widetilde{\mathcal{A}}$ and $\widetilde{\mathcal{B}}$ via a series of calls to a micro-kernel (see Fig. 4, top). The micro-kernel represents a wide inner product of a $m_R \times k_C$ sub-matrix of $\widetilde{\mathcal{A}}$ with a $n_R \times k_C$ sub-matrix of $\widetilde{\mathcal{B}}$ as a series of outer products, of size $m_R \times n_R$, for which the resulting auxiliary array $\widetilde{\mathcal{C}}$ can be entirely kept in registers (e.g., $m_R = 24$, $n_R = 4$ for single-precision and $m_R = 12$, $n_R = 4$ for double-precision calculations on the Intel Haswell microarchitecture).

While $\widetilde{\mathcal{A}}$ and $\widetilde{\mathcal{B}}$ are packed, $\widehat{\mathcal{C}}$ is not; in contrast to a high-performance GEMM (see Sec. 3.2) $\widehat{\mathcal{C}}$ can be of arbitrary dimensionality. We therefore represent this sub-tensor by $\widehat{\mathcal{C}}(\widetilde{m}_1, \widetilde{n}_1, \widetilde{m}_2, \widetilde{n}_2)$ such that its dimensionality is not fixed and that its dependencies on the free indices of $\widetilde{\mathcal{A}}$ and $\widetilde{\mathcal{B}}$ are explicit. The arrow labeled with "update" illustrates that the packed sub-tensor $\widetilde{C}_{\widetilde{m}_1,\widetilde{n}_1}$ updates the appropriate portion of the sub-tensor $\widehat{C}(\widetilde{m}_1, \widetilde{n}_1, \widetilde{m}_2, \widetilde{n}_2)$; we refer to this process as "unpacking". As we will discuss in the next section, this update—in the general case—denotes a tensor transposition (see Sec. 3.2).

### 3.2. Packing

The key observation is that the packed arrays $\widetilde{\mathcal{A}}$, $\widetilde{\mathcal{B}}$, and $\widetilde{\mathcal{C}}$ can be logically described as three-, three-, and two-dimensional tensors, respectively (i.e., $\widetilde{\mathcal{A}}_{\widetilde{m}_1,\widetilde{k}_1,\widetilde{m}_2}$, $\widetilde{\mathcal{B}}_{\widetilde{n}_1,\widetilde{k}_1,\widetilde{n}_2}$, and $\widetilde{\mathcal{C}}_{\widetilde{m}_1,\widetilde{n}_1}$)—as opposed to two-dimensional arrays. Representing these auxiliary ar-rays as packed sub-tensors enables us to express the (un)packing routines in terms of tensor transpositions. This "transformation" is advantageous because tensor transpositions can be fully vectorized and typically attain close to a system's peak memory bandwidth [Springer et al. 2017; Springer et al. 2016a; Springer et al. 2016b].

While blocking enables us to effectively reduce tensor contractions to three nested loops around a macro-kernel, packing the data into contiguous arrays avoids severe conflict misses which occur due to the limited associativity of modern caches; the packed format also increases the spacial locality of the sub-tensors, reducing TLB misses within the macro-kernel. In sharp contrast to matrix-matrix multiplications, which only deal with 2D sub-matrices, the packing routines for GETT might require to fetch data from arbitrary dimensional sub-tensors and pack it into contiguous, auxiliary arrays.

To pack parts of $\mathcal{A}$, $\mathcal{B}$ and $\mathcal{C}$, one has to: (1) identify sub-tensors $\widehat{\mathcal{A}}$, $\widehat{\mathcal{B}}$ and $\widehat{\mathcal{C}}$ (of appropriate size) from the original tensors $\mathcal{A}$, $\mathcal{B}$, $\mathcal{C}$ and then (2) transpose these non-contiguous sub-tensor $\widehat{\mathcal{A}}$, $\widehat{\mathcal{B}}$, $\widehat{\mathcal{C}}$ into $\widetilde{\mathcal{A}}$, $\widetilde{\mathcal{B}}$ and $\widetilde{\mathcal{C}}$, respectively.[13] Furthermore, in order to express these packing routines as tensor transpositions, one has to ensure that the total number of elements, as well as the dimensionality of the sub-tensors and their packed counterparts are identical. As an example, we now discuss the identification and the packing of a sub-tensor for $\widehat{\mathcal{A}}$; the process of packing $\widehat{\mathcal{B}}$, and that of unpacking $\widetilde{\mathcal{C}}$ work similarly.

*3.2.1. Identify sub-tensor.* Given the desired blocking sizes $m_C$ and $k_C$,[14] fitting the cache capacity of the underlying processor, we seek to identify two subsets $I_{\widehat{m}} \subseteq I_m$ and $I_{\widehat{k}} \subseteq I_k$ with

$$S_{I_{\widehat{m}}} = m_C, \tag{5}$$
$$S_{I_{\widehat{k}}} = k_C \tag{6}$$

to form the sub-tensor $\widehat{\mathcal{A}}_{\Pi^{\widehat{\mathcal{A}}}(I_{\widehat{m}} \cup I_{\widehat{k}})}$. $\Pi^{\widehat{\mathcal{A}}}$ denotes almost the same permutation as the original permutation $\Pi^{\mathcal{A}}$ but with all indices $i \in (I_m \cup I_k) \setminus (I_{\widehat{m}} \cup I_{\widehat{k}})$ removed; more precisely, the order of the indices remains unchanged. For instance, given the original index sets $I_m = \{m_1, m_2\}$, $I_k = \{k_1, k_2\}$ and the index sets of the packed tensors, of appropriate size, $I_{\widehat{m}} = \{m_1\}$ and $I_{\widehat{k}} = \{k_2\}$ as well as the permutation $\Pi^{\mathcal{A}}(I_m \cup I_k) = (m_1, k_1, m_2, k_2)$, then $\Pi^{\widehat{\mathcal{A}}}$ is expressed as: $\Pi^{\widehat{\mathcal{A}}}(I_{\widehat{m}} \cup I_{\widehat{k}}) = (m_1, k_2)$.

In addition to the constraints (5)-(6), we also request that the stride-one index of $\mathcal{A}$ is part of either $I_{\widehat{m}}$ or $I_{\widehat{k}}$. Even though this condition is not required from a correctness perspective, it is necessary for high performance. The fact that the stride-one index is part of $\widehat{\mathcal{A}}$ enables vectorized loads and stores throughout the tensor transpositions; thus, yielding more efficient packing routines. Fig. 5a illustrates two examples of "suitable" sub-tensors which preservers the stride-one index; Fig. 5b, on the other hand, depicts a "non suitable" sub-tensor for which the stride-one index is not preserved.

Depending on the size and shape of $\mathcal{A}$, it might not be immediately possible to identify suitable subsets $I_{\widehat{m}}$ and $I_{\widehat{k}}$ that respect the aforementioned constraints (see Examples below). Thus, in order to find such subsets we might need to reinterpret $\mathcal{A}_{\Pi^{\mathcal{A}}(I_m \cup I_k)}$ as a different tensor $\mathcal{A}'_{\Pi^{\mathcal{A}'}(I'_m \cup I'_k)}$ of the same size (i.e., $S_{I_m} = S_{I_{m'}}$ and $S_{I_k} = S_{I_{k'}}$) but of a different dimensionality (i.e., $|I_m| \neq |I_{m'}|$ or $|I_k| \neq |I_{k'}|$). It is important to mention that the memory-layout of the tensor in question does not change; the shape of

---

[13]The direction for the update to $\widehat{\mathcal{C}}$ is reversed; $\widetilde{\mathcal{C}}$ is unpacked into $\widehat{\mathcal{C}}$.
[14]To simplify, we require $S_{I_m}$ to be a multiple of $m_R$. Lifting this constraint is a future task.

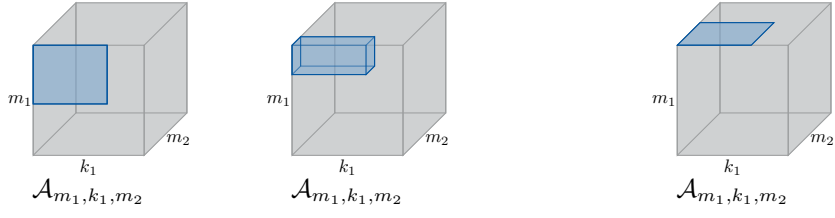

(a) Good: Stride-one index preserved.    (b) Bad: Stride-one index not preserved.

Fig. 5: Sub-tensors. Volume of the highlighted sub-tensors is not to scale.

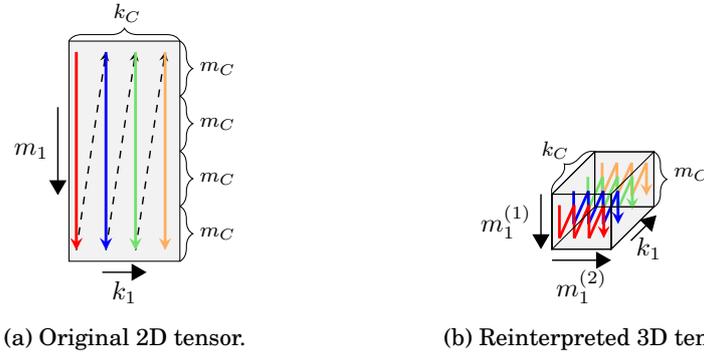

(a) Original 2D tensor.    (b) Reinterpreted 3D tensor.

Fig. 6: Reinterpreting a two-dimensional tensor $\mathcal{A}_{m_1,k_1}$ as a three-dimensional tensor $\mathcal{A}'_{m_1^{(1)},m_1^{(2)},k_1}$ of the same size. The colored arrows depict the memory layout.

the tensor is merely reinterpreted. Hence, this change of perspective does not require any memory accesses and thus happens entirely at compile-time. To give the reader a better intuition about this process, we look at two examples which require $\mathcal{A}$ to be reinterpreted.

***Example 1:*** Let $\mathcal{A}_{m_1,k_1}$ be a two dimensional tensor with $S_{m_1} = 4m_C$ and $S_{k_1} = k_C$ (see Fig. 6a), our objective is to identify a sub-tensor $\widehat{\mathcal{A}}_{\Pi_{\widehat{\mathcal{A}}}(I_{\widehat{m}} \cup I_{\widehat{k}})}$ of size $m_C \times k_C$ with $S_{I_{\widehat{m}}} = m_C$ and $S_{I_{\widehat{k}}} = k_C$. This can be achieved by splitting the index $m_1$ into $m_1^{(1)}$ and $m_1^{(2)}$ such that the combined size of $m_1^{(1)}$ and $m_1^{(2)}$ remains the same as the size of the original index $m_1$ (i.e., $S_{m_1} = S_{m_1^{(1)}} \times S_{m_1^{(2)}}$)—effectively reinterpreting the 2D tensor $\mathcal{A}_{m_1,k_1}$ as the 3D tensor $\mathcal{A}'_{m_1^{(1)},m_1^{(2)},k_1}$ (see Fig. 6b). Once $\mathcal{A}$ has been reinterpreted as a 3D tensor, one can define the subsets $I_{\widehat{m}} := \{m_1^{(1)}\}$ and $I_{\widehat{k}} := \{k_1\}$. Thus we identified the desired sub-tensor $\widehat{\mathcal{A}}_{m_1^{(1)},k_1}$ which complies to Eq. (5)-(6).

***Example 2:*** We now focus on a more involved example. Let $\mathcal{A}_{m_1,k_1,m_2}$ be a 3D tensor for which neither $S_{m_1}$ nor $S_{m_2}$ is a multiple of $m_C$, while their product $S_{m_1} \times S_{m_2}$ is. Moreover, $S_{k_1} = k_C$ is the same as in the previous example. To find a subset $I_{\widehat{m}}$ with a total size of $m_C$ elements, $\mathcal{A}$ can be reinterpreted as a 5D tensor $\mathcal{A}'_{m_1^{(1)},m_1^{(2)},k_1,m_2^{(1)},m_2^{(2)}}$ and assign $I_{\widehat{m}} := \{m_1^{(1)}, m_2^{(1)}\}$ with $S_{I_{\widehat{m}}} = m_C$. Similarly to the previous example, we enforce that the total size of the split indices remains the same (i.e., $S_{m_1} = S_{m_1^{(1)}} S_{m_1^{(2)}}$ and $S_{m_2} = S_{m_2^{(1)}} \times S_{m_2^{(2)}}$). Thus, the desired sub-tensor $\widehat{\mathcal{A}}_{m_1^{(1)},k_1,m_2^{(1)}}$ can be identified out of the 5D tensor $\mathcal{A}'_{m_1^{(1)},m_1^{(2)},k_1,m_2^{(1)},m_2^{(2)}}$; again, the order of the indices is not allowed to change.

The astute reader might have noticed that the splitting of any index has to be based on the prime factors of its size. While a detailed discussion of this mechanism is beyond the scope of this paper, we point out that our GETT implementation splits all indices based on their prime factors, and selects suitable subsets $I_{\widehat{m}}$, $I_{\widehat{n}}$ and $I_{\widehat{k}}$ (of appropriate size) such that the stride-one indices of $\mathcal{A}, \mathcal{B}, \mathcal{C}$ are preserved in the sub-tensors $\widehat{\mathcal{A}}, \widehat{\mathcal{B}}$ and $\widehat{\mathcal{C}}$.

Since multiple index sets exist which fulfill the aforementioned conditions, we currently explore all candidates (see Sec. 3.3) and use a performance model to rank them automatically (see Sec. 3.4).

*3.2.2. Packing via tensor transpositions.* While the memory layout of the packed sub-tensors $\widetilde{\mathcal{A}}, \widetilde{\mathcal{B}}$ and $\widetilde{\mathcal{C}}$ (see Fig. 4) as well as the sizes of their indices are always fixed (i.e., $S_{\widetilde{m}_1} = m_R$, $S_{\widetilde{n}_1} = n_R$, $S_{\widetilde{m}_2} = \frac{m_C}{S_{\widetilde{m}_1}}$, $S_{\widetilde{n}_2} = \frac{m_C}{S_{\widetilde{m}_1}}$ and $S_{\widetilde{k}_1} = k_C$), the dimensionality of the non-packed sub-tensors $\widehat{\mathcal{A}}, \widehat{\mathcal{B}}$ and $\widehat{\mathcal{C}}$, on the other hand, can vary from a TC to another. Thus, we need to define a reliable mapping between these tensors for an arbitrary TC.

Recall that the index sets of the free indices $I_{\widehat{m}}$ and the contracted indices $I_{\widehat{k}}$ of the non-packed sub-tensor $\widehat{\mathcal{A}}$ already have the appropriate—and fixed—size (i.e., $S_{I_{\widehat{m}}} = m_C$ and $S_{I_{\widehat{k}}} = k_C$). Moreover, the value of $m_C$ is chosen such that it is a multiple of $m_R$. Similarly to the previous section we have to identify two non-overlapping subsets $I_{\widehat{m}}^0, I_{\widehat{m}}^1 \subset I_{\widehat{m}}$ such that

$$S_{I_{\widehat{m}}^0} = m_R, \tag{7}$$

$$S_{I_{\widehat{m}}^1} = \frac{S_{I_{\widehat{m}}}}{S_{I_{\widehat{m}}^0}}. \tag{8}$$

Since those subsets might not exists, the shape of $\widehat{\mathcal{A}}_{\Pi^{\widehat{A}}(I_{\widehat{m}} \cup I_{\widehat{k}})}$ has to be reinterpreted once more to form $\check{A}_{\Pi^{\check{A}}(I_{\widehat{m}}^0 \cup I_{\widehat{k}} \cup I_{\widehat{m}}^1)}$ via the same mechanism discussed in the previous section.

Reinterpreting the shape of $\widehat{\mathcal{B}}$ and $\widehat{\mathcal{C}}$ in a similar fashion yields suitable sub-tensors $\check{B}$ and $\check{C}$. We are finally able to encode the packing of the non-packed sub-tensors $\check{A}, \check{B}, \check{C}$ into their packed counterparts $\widetilde{\mathcal{A}}, \widetilde{\mathcal{B}}, \widetilde{\mathcal{C}}$ as tensor transpositions of the form:

$$\widetilde{\mathcal{A}}_{I_{\widehat{m}}^0, I_{\widehat{k}}, I_{\widehat{m}}^1} \leftarrow \check{A}_{\Pi^{\check{A}}(I_{\widehat{m}}^0 \cup I_{\widehat{k}} \cup I_{\widehat{m}}^1)}, \tag{9}$$

$$\widetilde{\mathcal{B}}_{I_{\widehat{n}}^0, I_{\widehat{k}}, I_{\widehat{n}}^1} \leftarrow \check{B}_{\Pi^{\check{B}}(I_{\widehat{n}}^0 \cup I_{\widehat{k}} \cup I_{\widehat{n}}^1)}, \tag{10}$$

$$\check{C}_{\Pi^{\check{C}}(I_{\widehat{m}}^0 \cup I_{\widehat{n}}^0)} \leftarrow \alpha \times \widetilde{\mathcal{C}}_{I_{\widehat{m}}^0, I_{\widehat{n}}^0} + \beta \times \check{C}_{\Pi^{\check{C}}(I_{\widehat{m}}^0 \cup I_{\widehat{n}}^0)}. \tag{11}$$

***Example.*** Given the sub-tensor $\widehat{\mathcal{A}}_{k_1,m_1,k_2,m_2}$ with $k_C = S_{k_1} \times S_{k_2}$ and $S_{m_1} = a \times b$, $S_{m_2} = c \times d$, $a, b, c, d \in \mathbb{N}$ and $m_R = a \times c$. Then we can reinterpret $\widehat{\mathcal{A}}$ as $\check{A}_{k_1,m_1^{(1)},m_1^{(2)},k_2,m_2^{(1)},m_2^{(2)}}$ such that the $S_{m_1^{(1)}} = a$, $S_{m_1^{(2)}} = b$, $S_{m_2^{(1)}} = c$, $S_{m_2^{(2)}} = d$; resulting in the tensor transposition: $\widetilde{\mathcal{A}}_{m_1^{(1)},m_2^{(1)},k_1,k_2,m_1^{(2)},m_2^{(2)}} \leftarrow \check{A}_{k_1,m_1^{(1)},m_1^{(2)},k_2,m_2^{(1)},m_2^{(2)}}$. Notice that other permutations would also have been possible (e.g., $\widetilde{\mathcal{A}}_{m_1^{(1)}, m_2^{(1)}, k_2, k_1, m_2^{(2)}, m_1^{(2)}} \leftarrow \check{A}_{k_1, m_1^{(1)}, m_1^{(2)}, k_2, m_2^{(1)}, m_2^{(2)}}$).

### 3.3. Search Space

This section outlines the search space of viable implementations (henceforth called candidates) for any TC which is cast in terms of GETT. The combination of the different GEMM variants (see Table I), the choice of the blocking sizes (i.e., $m_C, n_C, k_C, m_R$ and $n_R$) as well as the permutations of free and contracted index sets (i.e., $I_m, I_n$ and $I_k$) constitutes a large search space of viable candidates. Instead of exhaustively testing all candidates (i.e., generating, compiling and timing.), we rank them according to an architecture-aware metric (see next section); thanks to this metric the search space is pruned significantly.

Depending on the size of the index sets $I_m, I_n$ and $I_k$, it might be useful to favor one of the three GEMM-variants (outlined in Table I of Sec. 2.1) over another. Hence, GETT examines all three variants for each TC.

Our current GETT implementation explores up to four different parameters for each $m_C, n_C$ and $k_C$, while keeping $m_R$ and $n_R$ fixed. More precisely, $m_R$ and $n_R$ are chosen such that the latencies of the fused-multiply-add (FMA) instructions within the micro-kernel are completely hidden in order to achieve peak floating-point performance [Low et al. 2015]. A future GETT implementation might also explore different $m_R, n_R$ values; this could be useful in the bandwidth-bound regime, where one would be willing to trade a less efficient micro-kernel for more efficient packing routines.

Finally, as we have already alluded to in the previous section, there are several ways to identify suitable sub-tensors out of $\mathcal{A}, \mathcal{B}$ and $\mathcal{C}$. We restrict the search (of viable sub-tensors) to those sub-tensors which preserve the stride-one index in the corresponding tensor $\mathcal{A}, \mathcal{B}$ or $\mathcal{C}$.

### 3.4. Performance Model

We model the runtime for any GETT candidate depending on its selected GEMM-variant, the sizes of the packed sub-tensors, and the permutations of the required packing routines. This process is broken down into two stages: (1) estimate the time for the packing routines, and (2) estimate the time for the macro-kernel. The total time for the current candidate is estimated as the sum of (1) and (2).

For the sake of simplicity, let us again focus on the example of a matrix-panel multiplication (i.e., the algorithm outlined in Listing 4). The packing of $\widehat{\mathcal{B}}$ into $\widetilde{\mathcal{B}}$ reads and writes a total of

$$\text{Data}_\mathcal{B} = 2 \times S_{I_k} \times S_{I_n} \times \text{sizeof(floatType)} \qquad (12)$$

bytes, over the course of the entire tensor contraction. This operation is clearly bandwidth-bound; thus, we can estimate the time that it takes to pack $\widehat{\mathcal{B}}$ by

$$\text{Time}_\mathcal{B} = \frac{\text{Data}_\mathcal{B}}{BW \times p_D}, \qquad (13)$$

where $BW$ represents the system's SAXPY-Bandwidth[15] and $p_D \in (0, 1]$ is a penalty which favors some permutations over others. For instance, the performance model slightly prefers small-dimensional transpositions over high-dimensional ones, since the former typically exhibit a more regular memory access pattern (i.e., $p_D$ slightly de-creases with increasing dimensionality $d$, $p_D = 1.0 - (d-1) \times 0.015$). Moreover, previous research [Springer et al. 2016a; Lyakh 2015] suggests that transpositions for which the stride-one index does not change (e.g., $\widetilde{A}_{i,j,k} \leftarrow \widehat{A}_{i,k,j}, \widetilde{A}_{i,j,k,l} \leftarrow \widehat{A}_{i,l,k,j}$) are more efficient than others. Hence, we penalize those transpositions for which the stride-one index changes by decreasing $p_D$ by an additional 30% (i.e., $p_d \leftarrow p_d \times 0.7$). A similar analysis is carried out for $\text{Time}_\mathcal{A}$ and $\text{Time}_\mathcal{C}$, the only difference being that $A$ and $C$ will be (un)packed $\frac{S_{I_n}}{n_C}$ and $\frac{S_{I_k}}{k_C}$ many times, respectively.

---

[15] As defined by the BLAS, SAXPY is the single precision vector-vector addition ($\mathbf{y} \leftarrow \alpha \times \mathbf{x} + \mathbf{y}, \alpha \in \mathbb{R}, \mathbf{x}, \mathbf{y} \in \mathbb{R}^n$).

Finally, we estimate the time of the macro-kernel according to

$$\text{Time}_{\text{macro}} = \frac{2 \times S_{I_m} \times S_{I_n} \times S_{I_k}}{FP^{\text{peak}} \times p_F}, \tag{14}$$

where $FP^{\text{peak}}$ denotes the system's theoretical peak floating-point performance and $p_F \in (0, 1]$ is again a penalty. We penalize the performance of the micro-kernel by 30% (i.e., $p_F \leftarrow p_F \times 0.7$), whenever one of the following conditions is violated: 1) $\widetilde{\mathcal{B}} \in \mathbb{R}^{n_C \times k_C}$ fits into the L3 cache, 2) $\widetilde{\mathcal{A}} \in \mathbb{R}^{m_C \times k_C}$ alongside two $n_R \times k_C$ sub-blocks of $\widetilde{\mathcal{B}}$ fit into L2 cache ($\mathcal{B}$ is streamed through L2), and 3) a $n_R \times k_C$ sub-block of $\widetilde{\mathcal{B}}$ alongside a two $m_R \times k_C$ sub-blocks of $\widetilde{\mathcal{A}}$ fit simultaneously fit into the L1 cache ($\mathcal{A}$ is streamed through L1).

The total estimated time for a candidate is then given by $\text{Time}_{\text{macro}} + \text{Time}_\mathcal{A} + \text{Time}_\mathcal{B} + \text{Time}_\mathcal{C}$; while this is still a rather rough model, it works well in practice (see Sec. 7).

## 4. TRANSPOSE-TRANSPOSE-GEMM-TRANSPOSE

In this section we discuss the principles of our TTGT code generator. The idea behind TTGT is to rearrange (transpose) the elements of a tensor such that it can be logically interpreted as a matrix. This initial step, also referred to as flattening, makes it possible to compute the contraction via an ordinary matrix-matrix multiplication, thus, exploiting GEMM's high efficiency. Depending on the actual contraction, the resulting matrix has to be folded (transposed) back.

```
1   candidates = [] // this set keeps track of the candidates
2   I_m = A.indices ∩ C.indices; // free indices from A
3   I_n = B.indices ∩ C.indices; // free indices from B
4   I_k = A.indices ∩ B.indices; // contracted indices
5   for all I'_m in permutations(I_m):
6       for all I'_n in permutations(I_n):
7           for all I'_k in permutations(I_k):
8               // Candidate 1: A_{I'_m, I'_k}, B_{I'_n, I'_k}
9               candidates.append( [
10                  A_{I'_m, I'_k} ← A_{Π_A(I_m ∪ I_k)},   // unfold A
11                  B_{I'_n, I'_k} ← B_{Π_B(I_n ∪ I_k)},   // unfold B
12                  X_{I'_m, I'_n} ← A_{I'_m, I'_k} × B^T_{I'_n, I'_k},   // GEMM
13                  C_{Π_C(I_m ∪ I_n)} ← X_{I'_m, I'_n} ] );   // fold X
14              // Candidate 2: A_{I'_m, I'_k}, B_{I'_k, I'_n} (omitted)
15              // Candidate 3: A_{I'_k, I'_n}, B_{I'_n, I'_k} (omitted)
16              // Candidate 4: A_{I'_k, I'_n}, B_{I'_k, I'_n} (omitted)
17              // Candidate 5: B_{I'_n, I'_k}, A_{I'_m, I'_k} (omitted)
18              // Candidate 6: B_{I'_n, I'_k}, A_{I'_k, I'_m} (omitted)
19              // Candidate 7: B_{I'_k, I'_n}, A_{I'_m, I'_k} (omitted)
20              // Candidate 8: B_{I'_k, I'_n}, A_{I'_k, I'_m} (omitted)
21  return select( candidates );  // pick most promising candidate
```

Listing 5: Transpose-Transpose-GEMM-Transpose code generator.

Listing 5 outlines the pseudo code for the TTGT code generator. In Lines 2–4, the free indices from $\mathcal{A}$ (i.e., $I_m$) and $\mathcal{B}$ (i.e., $I_n$) as well as the contracted indices (i.e., $I_k$) are extracted. Next, all permutations of these index sets are considered (Lines 5–7). Once the permutations of the index sets are fixed, eight different TTGT candidates are added to the set candidates (Lines 9–20). A candidate comprises up to three tensor transpositions (Lines 10, 11, 13) as well as one matrix-matrix multiplication (Line 12); notice that, depending on the actual tensor contraction, some (or even all) of these transpositions might be redundant. The most promising candidate is then chosen according to a metric that minimizes the "(un)folding overhead", i.e., it minimizes the combined size of the tensors involved in the transpositions in Lines 10, 11, and 13. Thus, our TTGT implementation does not require any search but solely relies on the heuristic to pick the most promising candidate automatically. Similarly to the performance model discussed in Sec. 3.4 for the GETT approach, this metric also favours those transpositions for which the stride-one index remains unchanged.

**Example.** Let us consider the contraction $\mathcal{C}_{a,b,c} = \mathcal{A}_{a,c,k}\mathcal{B}_{b,k}$; in this case, neither $\mathcal{A}$ nor $\mathcal{B}$ need to be transposed. An exemplary implementation following the TTGT approach for this TC is shown in Listing 6.

```
1  X_{(a,c),b} ← A_{(a,c),k} × B^T_{b,k}  // contract A and B via a GEMM
2  C_{a,b,c} ← X_{a,c,b}  // fold X
```

Listing 6: TTGT pseudo code for $\mathcal{C}_{a,b,c} = \mathcal{A}_{a,c,k}\mathcal{B}_{b,k}$. The indices $a$ and $c$ have been reinterpreted as a "super-index" $(a, c)$.

While TTGT can perform very well in the compute-bound regime, it performs poorly for bandwidth-bound TCs. This suboptimal behaviour is due to its two major disadvantages: First, the (un)folded tensors require additional memory; second, each element (of a transposed tensors) needs to be loaded at least twice, effectively doubling the bandwidth requirement.

## 5. LOOPS-OVER-GEMM

This section outlines the code generation process based on the LoG approach. The strategy adopted in Listing 7 is to keep the stride-one indices as part of the looped-over GEMM; these indices are referred to as requiredIndices (Line 2). If the stride-one index of $\mathcal{C}$ belongs to $\mathcal{B}$, then LoG interchanges $\mathcal{A}$ and $\mathcal{B}$ (Lines 3–4). The free indices from $\mathcal{A}$ and $\mathcal{B}$ as well as the contracted indices are identified in Lines 5–7. An LoG implementation is only considered if none of the conditions in Lines 8–11 are violated; these conditions ensure that the 2D slices can be processed via an ordinary GEMM with strided accesses in only one dimension as opposed to strides in both dimension which results in suboptimal memory accesses.[16]

---

[16]To our knowledge, BLIS [Van Zee and van de Geijn 2015] is the only high-performance BLAS which supports strides in both matrix dimensions.

```
1   candidates = [] // this set keeps track of the candidates
2   requiredIndices = set( A.indices[0], B.indices[0], C.indices[0] );
3   if( C.indices[0] ∈ B.indices )
4       swap( A, B );
5   I_m = A.indices ∩ C.indices; // free indices from A
6   I_n = B.indices ∩ C.indices; // free indices from B
7   I_k = A.indices ∩ B.indices; // contracted indices
8   if( |requiredIndices ∩ I_m| ≥ 2 or
9       |requiredIndices ∩ I_n| ≥ 2 or
10      |requiredIndices ∩ I_k| ≥ 2 )
11      return -1; // LoG not possible
12  while( moreCandidatesPossible() )
13      mIdx = I_m ∩ requiredIndices;
14      if( mIdx == None )
15          mIdx = pickOneIndexFrom( I_m );
16      nIdx = I_n ∩ requiredIndices;
17      if( nIdx == None )
18          nIdx = pickOneIndexFrom( I_n );
19      kIdx = I_k ∩ requiredIndices;
20      if( kIdx == None )
21          kIdx = pickOneIndexFrom( I_k );
22      loopIndices = (I_m ∪ I_n ∪ I_k) \ {mIdx, nIdx, kIdx};
23      for all loopOrder in perm(loopIndices)
24          // all loop orders are viable
25          candidates.append( Candidate(mIdx, nIdx, kIdx, loopOrder) );
26
27  return select( candidates ); // pick most promising candidate
```

Listing 7: Loops-over-GEMM code generator.

In Lines 12–25, all possible candidates are generated, and then stored to the `candidates` list. Lines 13–21 determine the m-, n- and k-index of the GEMM call; notice that multiple candidates are possible if `mIdx`, `nIdx` or `kIdx` are not already covered by `requiredIndices`. The remaining indices constitute the list of looped-over indices (Line 22). All permutations of the `loopIndices` are considered (Lines 23–25) and the resulting candidates are appended to `candidates`. Finally, in Line 27 the most promising candidate is selected according to a metric that ranks the candidate based on the flop-count of the GEMM (i.e., $2 \times S_{\texttt{mIdx}} \times S_{\texttt{nIdx}} \times S_{\texttt{kIdx}}$); the rationale being that larger GEMMs typically exhibit higher performance.[17] A possible extension to this model could be to also account for the estimated efficiency of the corresponding GEMM call [Peise and Bientinesi 2012].

---

[17]Caveat: this metric is fooled by skewed GEMMs for which at least one index is small but the product of all indices is large; those GEMMs typically also result in poor performance.

```c
void loops_over_gemm(const float *A, const float *B, float *C,
                     float alpha, float beta)
{
    const int i_UP=576, m_UP=24, c_UP=24, k_UP=24, a_UP=576;
    const int m_  = a_UP;
    const int n_  = i_UP;
    const int k_  = m_UP;
    const int lda_ = m_UP * k_UP;
    const int ldb_ = i_UP;
    const int ldc_ = a_UP * c_UP;
    // loop over free indices
    for(int c=0; c < c_UP; c++)
       for(int k=0; k < k_UP; k++){
          sgemm_("T", "T", &m_, &n_, &k_, &alpha,
             &B[k * (m_UP)], &lda_,
             &A[c * (i_UP*m_UP )], &ldb_, &beta,
             &C[c * (a_UP) + k * (a_UP*c_UP*i_UP)], &ldc_);
       }
}
```

Listing 8: Generated LoG code for $C_{a,b,c,i,j,k} = A_{i,j,m,c} B_{m,k,a,b}$; each index is of size $24$.

Listing 8 shows the LoG code generated for an exemplary tensor contraction of the form $C_{a,b,c,i,j,k} = A_{i,j,m,c} \times B_{m,k,a,b}$, where each index is of size $24$. This candidate loops over the $k$ and $c$ index (Lines 10, 11) and contracts the $m$ index via a GEMM (Lines 13–16). It is important to notice that the indices $a, b$ and $i, j$ are merged/flattened into super-indices $(i, j)$ and $(a, b)$, respectively;[18] hence, one can interpret the GEMM as $C_{(a,b),(i,j)} = B^T_{m,(a,b)} \times A^T_{(i,j),m}$, where $^T$ denotes a matrix-transpose.

## 6. TENSOR CONTRACTION CODE GENERATOR

```
C[a,b,i,j] = A[i,m,a] * B[m,j,b]
a = 24
b = 24
i = 24
j = 24
m = 24
```

Listing 9: Content of an exemplary TCCG input file.

Having presented three different approaches (GETT, TTGT and LoG), we now describe the **T**ensor **C**ontraction **C**ode **G**enerator (TCCG), a unified code generator for TCs written in Python. The input to TCCG is a contraction with the size of each index (see Listing 9); the output is high-performance C++ code. TCCG is publicly available at www.github.com/springer13/tccg.

A schematic overview of TCCG is presented in Fig. 7. Before starting the code-generation process, TCCG merges consecutive indices in the tensors to super-indices (Stage 1). Still as a preprocessing stage, a local SQL database of known implementations is queried to check if a solution for the specific contraction (and size) already exists; if so, no generation takes place, and the previous implementation is returned. Otherwise, Stage 2 takes place: TCCG maintains a list of promising candidates throughout the code-generation process;[19] candidates are generated based on the GETT, LoG and TTGT approaches. In Stage 3, the candidates are "evaluated" according to a performance model; a candidate is stored to the internal list if the cost of the current candidate is smaller than the cost $\kappa$ of the worst candidate within the list. This process continues until all candidates have been considered. The most promising candidates are then compiled (4) and timed (5). Finally, in Stage 6, the fastest candidate is stored to the SQL database, and the corresponding C++ code is generated.

---

[18]Two indices $i$ and $j$ which are appear consecutively as $i, j$ in two tensors are merged/flattened into one super-index $(ij)$ of the same size (i.e., $S_{ij} = S_i S_j$).
[19]The capacity of this list is user-defined.

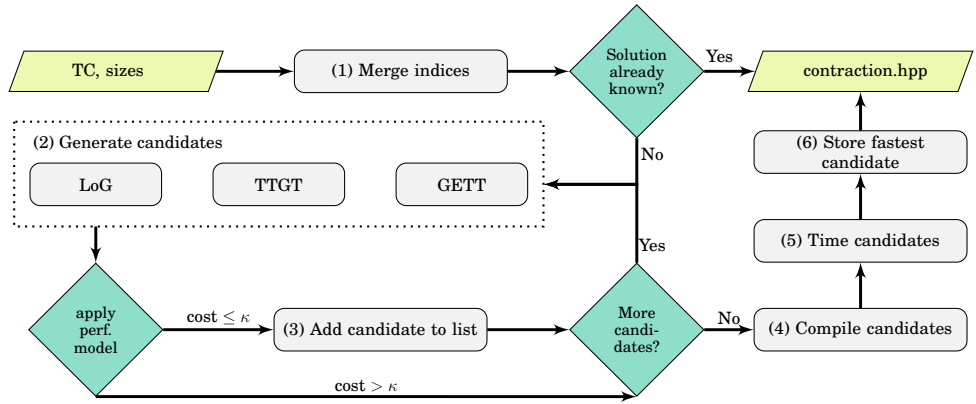

Fig. 7: Schematic overview of TCCG. $\kappa$ denotes the cost of the worst candidate within the maintained list of candidates.

| Argument | Description |
| --- | --- |
| `--floatType=[s,d]` | data type |
| `--maxWorkspace=<value>` | maximum auxiliary workspace in GB |
| `--maxImplementations=<value>` | maximum #implementations for GETT |
| `--arch=[avx2,avx512,cuda]` | selected architecture |
| `--numThreads=<value>` | number of threads |
| `--help` | prints all available command line options |

Table II: Subset of TCCG's command line arguments

TCCG offers users the possibility to influence TCCG's code-generation process via command-line options. For instance, users can either restrict the capacity of TCCG's internal list of candidates (via `--maxImplementations`) or limit the amount of auxiliary workspace that TTGT is allowed to use (via `--maxWorkspace`). The current implementation of TCCG supports both single- and double-precision calculations. Moreover, both TTGT as well as LoG are also available for CUDA-enabled GPUs.

## 7. PERFORMANCE RESULTS

In this section, we evaluate the performance of GETT, LoG, and TTGT separately, and then collectively, by means of TCCG, on a single core of an Intel *Xeon E5-2680 v3* CPU based on the Intel *Haswell* microarchitecture. ECC is enabled, and both *Intel Speedstep* and *Intel TurboBoost* are disabled for all measurements. The C++ compiler of choice is Intel's *icpc 16.0.1 20151021* with flags `-O3 -xhost -mkl`.

We report the minimum runtime over three runs, while clearing the caches in between each run (cold data). The correctness of the generated code is checked against a naive loop-based implementation which is similar to the naive tensor-tensor contraction outlined in Listing 3; this implementation serves as a lower bound on perfor-mance (we refer to this as "reference"). Additionally, we also report the performance of a GEMM of the same size (i.e., $m = S_{I_m}$, $n = S_{I_n}$, $k = S_{I_k}$) as the given TC (i.e., $A \in \mathbb{R}^{S_{I_m} \times S_{I_k}}, B \in \mathbb{R}^{S_{I_k} \times S_{I_n}}, C \in \mathbb{R}^{S_{I_m} \times S_{I_n}}$); the GEMM mimics the TTGT approach, but omits the explicit transpositions and thus yields incorrect results—it can be thought of as an upper bound for performance.

### 7.1. Benchmark

To facilitate performance comparisons of different approaches to tensor contractions, we compiled a tensor contraction benchmark, containing a wide range of use cases collected from previous publications [Apra et al. 2014; Stock et al. 2012b; Baumgartner et al. 2005; Li et al. 2015]. The benchmark, publicly accessible at www.github.com/springer13/tccg, consists of 48 different contractions.

— [Apra et al. 2014]: 18 bandwidth-bound contractions encountered in the CCSD(T) method [Raghavachari et al. 1989]. The corresponding tensors are of high dimensionality (i.e., two 4D input tensors forming a 6D output tensor).
— [Stock et al. 2012b]: 19 contractions encountered in coupled-cluster methods, using 2D to 4D tensors.
— [Baumgartner et al. 2005]: 3 contractions "often used in quantum chemistry calculations to transform a set of two-electron integrals from an atomic orbital basis to a MO basis".
— [Li et al. 2015]: 8 contractions contracting a 3D, 4D, or 5D tensor with a 2D one (i.e., tensor-matrix multiplication).

The sizes of the indices are chosen such that they reproduce those from the respective publication (when disclosed). Moreover, the benchmark ensures that the total memory consumption, per TC, is at least 200 MiB, thus, significantly larger than the last level cache of our test system (the actual sizes can be found in Appendix A).

In the following sections, to focus the readers' attention on those contractions that exhibit different performance characteristics, we only report results for a subset of the 48 contractions. Results for the full benchmark are provided in Appendix A.

To simplify the presentation, we encode a contraction $\mathcal{C}_{\Pi^\mathcal{C}(I_m \cup I_n)} \leftarrow \mathcal{A}_{\Pi^\mathcal{A}(I_m \cup I_k)} \times \mathcal{B}_{\Pi^\mathcal{B}(I_n \cup I_k)}$ as $\Pi^\mathcal{C}(I_m \cup I_n)$-$\Pi^\mathcal{A}(I_m \cup I_k)$-$\Pi^\mathcal{B}(I_n \cup I_k)$; for instance, $\mathcal{C}_{ijkl} \leftarrow \mathcal{A}_{imjn} \mathcal{B}_{nlmk}$ is represented as $ijkl$-$imjn$-$nlmk$.

### 7.2. Performance evaluation

We start our performance evaluation[20] with a closer look at the TTGT approach, by comparing our TTGT-based implementation (denoted by TTGT) against the performance attained by the Cyclops Tensor Framework (CTF), Tensor Toolbox (TT) and libtensor.[21] CTF, TT as well as libtensor all implement the TTGT approach.

Fig. 8 reports the performance of both single and double precision TTGT-based implementations for the benchmark; the horizontal, black lines denote the performance attained by a GEMM of equivalent size. The top of the graph denotes the theoretical peak floating-point performance of the given CPU and the selected precision. The columns are sorted according to their single-precision GEMM performance. Thus, the TCs on the left of the plot are bandwidth-bound, while those on the right are compute-bound.

---

[20] All results are based on version *v0.1* of the tensor contraction benchmark.
[21] Both Tensor Toolbox and libtensor do not support single precision.

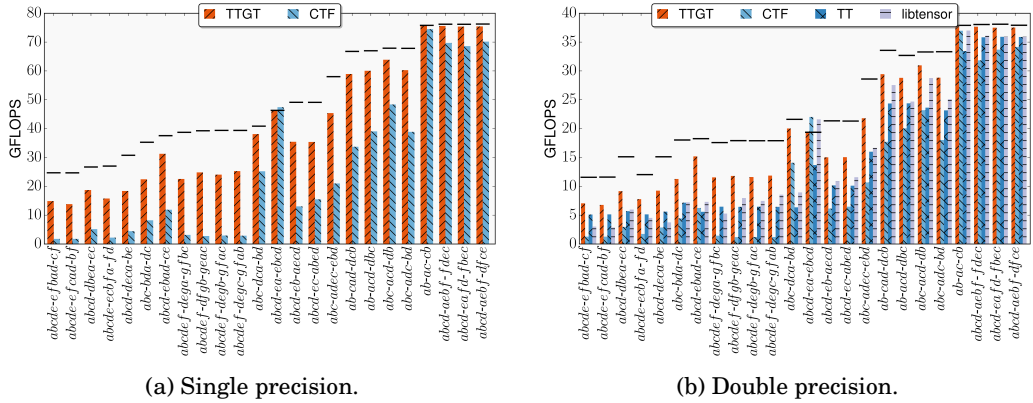

(a) Single precision.     (b) Double precision.

Fig. 8: TTGT performance. The horizontal black lines denote the performance of an equally-sized GEMM.

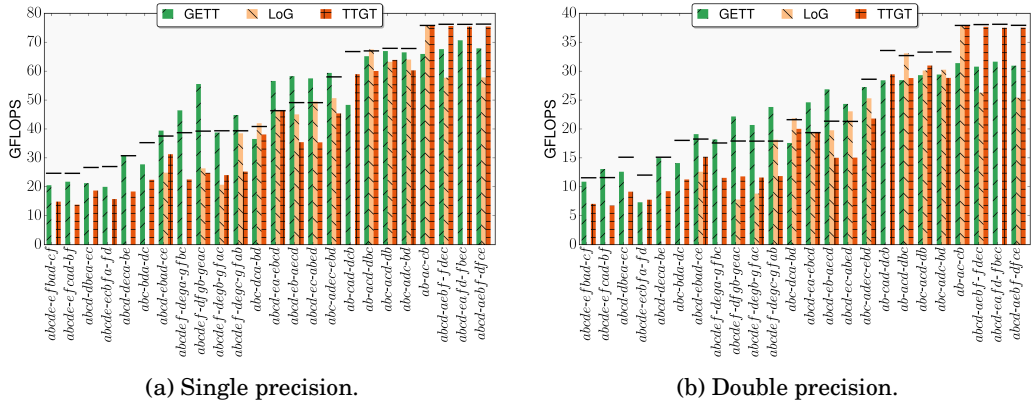

(a) Single precision.     (b) Double precision.

Fig. 9: GETT, LoG and TTGT performance. The horizontal black lines denote the performance of an equally-sized GEMM

We observe substantial speedups of our TTGT implementation over CTF, TT as well as libtensor across the entire benchmark.[22] As it is evident from these results, TTGT performs very well in the compute-bound regime–where the matrix-matrix multiplication dominates the execution time. However, it is also obvious that TTGT-based implementations do not yield optimal performance in the bandwidth-bound regime. The suboptimal performance can be attributed to the explicit transpositions, which accounts for pure overhead. More precisely, the overhead exhibited by these transpositions is shown by the difference between the solid black line and the reported TTGT performance; for instance, while bandwidth-bound TCs spent up to 50% of their runtime for (un)folding the tensors, compute-bound TCs only spent a negligible fraction of their runtime for these transpositions.

---

[22]With the exception of the *abcd-ea-ebcd* test case; this points to a performance bug within Intel's GEMM implementation; see https://software.intel.com/en-us/forums/intel-math-kernel-library/topic/636488 for further information.

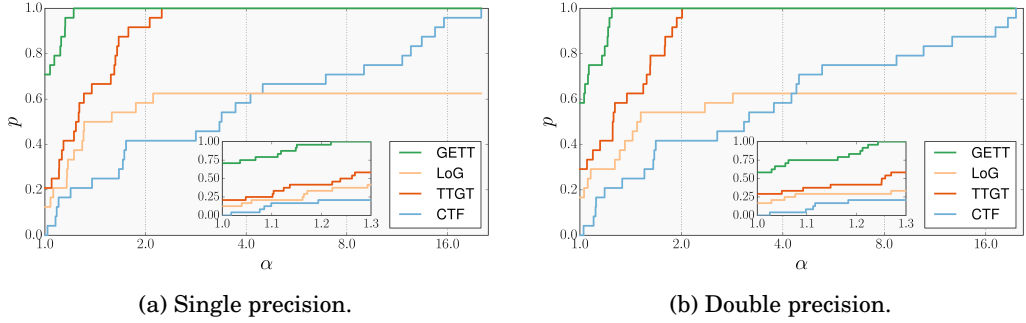

(a) Single precision.    (b) Double precision.

Fig. 10: Performance profiles. The inlets zoom into the range for which $\alpha \in [1, 1.3]$.

Fig. 9 continues our survey on different approaches to tensor contractions and presents the performance of GETT and LoG alongside TTGT. Looking at GETT's results (green bars), it is clear that GETT exhibits a significant speedup over the TTGT approach (red bars) in the bandwidth-bound regime (left side of the plot) while it does not quite reach TTGT's performance in the compute-bound regime (right side of the plot); we point out that GETT—in contrast to TTGT—does not require any auxiliary memory. For a number of test cases, GETT exceeds the reported GEMM performance, by up to $1.41\times$.[23] More precisely, GETT, on average, attains $98.1\%$ (minimum: $72.4\%$; maximum: $141.4\%$) and $97.0\%$ (minimum: $60.8\%$; maximum: $132.9\%$) of GEMM's performance across the benchmark for single precision (see Fig. 9a) and double precision (see Fig. 9b), respectively.

The LoG approach (orange bars in Fig. 9), on the other hand, shows variable performance across the benchmark. LoG improves the performance over TTGT for some TCs (e.g., $abcdef$-$degc$-$gfab$, $abcd$-$ec$-$abed$, $ab$-$acd$-$dbc$) but exhibits lower performance for others (e.g., $abcd$-$ebad$-$ce$, $abcd$-$aebf$-$fdec$). In certain situations (e.g., $abcde$-$efbad$-$cf$, $abcdef$-$dega$-$gfbc$), the approach is not applicable without an explicit transposition, or a GEMM implementation that allows strided memory accesses in multiple dimensions.

In Fig. 10, we combine the data from Fig. 8 and Fig. 9 and present it in the form of a performance profile [Dolan and Moré 2002], thus comparing GETT with CTF, LoG, and TTGT. For a given method $\mathcal{M}$, and a given point $\alpha$ on the $x$-axis, the corresponding value $p$ on the $y$-axis indicates the probability that $\mathcal{M}$ is at most a factor of $\alpha$ slower than the fastest of the four methods in question. Example: For $\alpha = 1.2$, TTGT has a $p$ value of about $0.4$; this means that in $40\%$ of the tests, TTGT is either the fastest approach, or within a factor of 1.2 from the fastest one. This plot makes it easy to draw some conclusions. (1) For about 70% (60%) of the test cases in single (double) precision, GETT is the fastest approach.[24] On about 30% (20%) of the test cases, TTGT is fastest; LoG is the method of choice for about 15% of cases, while CTF never is. (2) In those cases when GETT is not the fastest approach, it is never more than a factor of 1.22 (1.24) worse than the best solution.[25] TTGT and CTF are always within a factor of 2.2 (2.0) and 20.2 (19.7) from the best approach. (3) LoG's line plateaus at about $p = 0.6$; this indicates that for about 40% of the test cases the approach is not applicable.

---

[23]This points to suboptimal blocking choices within Intel's Math Kernel Library for the selected matrix-matrix multiplications.
[24]See the $p$-values for $\alpha = 1$.
[25]See the $\alpha$-values for which GETT reaches $p = 1$.

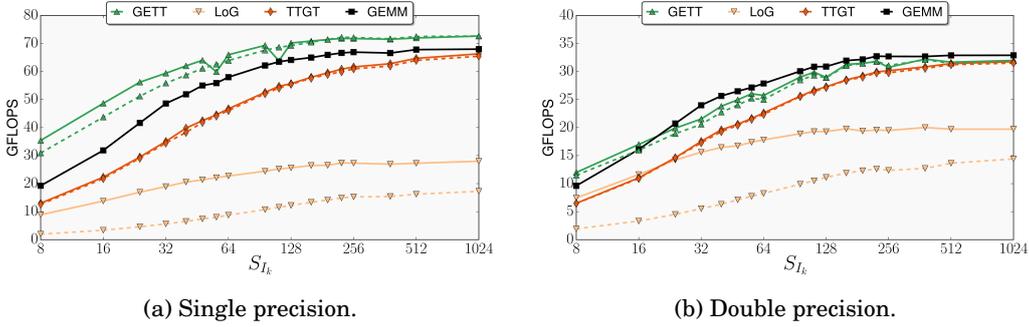

(a) Single precision.  (b) Double precision.

Fig. 11: Performance of GETT, LoG, TTGT and GEMM on the example of two similar tensor contractions with $S_{I_m} = S_{I_n} = 1152$. Solid line: $i_1ji_2$-$i_1ki_2$-$jk$, dashed line: $i_1j_1i_2j_2$-$i_1ki_2$-$j_1kj_2$.

Fig. 11 highlights the different performance characteristics of GETT, LoG and TTGT on the example of two similar tensor contractions. The sizes of the indices, for both TCs, are chosen such that $S_{I_m} = S_{I_n} = 1152$, while letting $S_{I_k}$ range from 8 to 1024. Hence, we effectively push the TC from the bandwidth-bound regime (left side of the plot) to compute-bound regime (right side of the plot). Analogously to our previous findings, one observes that GETT reaches up to 91.3% of peak performance in the compute-bound regime, and excels in the bandwidth-bound regime. Moreover, while LoG experiences a significant performance loss (compare solid and dashed lines), GETT and TTGT, on the other hand, are only marginally affected by the more complex tensor contraction $i_1j_1i_2j_2$-$i_1ki_2$-$j_1kj_2$. The performance loss of LoG can be explained by the fact that the size of the sub-matrices involved in the GEMM become smaller (i.e., $n = S_j$ for $i_1ji_2$-$i_1ki_2$-$jk$, but $n = \max(S_{j_1}, S_{j_2}) < S_j$ for $i_1j_1i_2j_2$-$i_1ki_2$-$j_1kj_2$); smaller sub-matrices result in lower arithmetic intensity and thus lower performance [Williams et al. 2009].[26] GETT and TTGT, on the other hand, retain the same arithmetic intensity and thus yield similar performance for both TCs.

Fig. 12 summarizes our performance discussion and combines the GETT, TTGT as well as LoG approaches (shown in green) to reflect the performance exhibited by TCCG. The code generated by TCCG is always on par with, and often significantly faster than the reference implementations (shown in red); notice, that the reference performance for single precision is lower than that of double precision due to the missing single precision support of Tensor Toolbox and libtensor (see Fig. 8). Compared to GEMM's performance, TCCG achieves for single and double precision a minimum/average/maximum of 74.0%/101.1%/141.4%% and 64.6%/101.8%/132.9%, respectively.

### 7.3. Evaluation of GETT Performance Model

Fig. 13 depicts the performance of the best $1, 4, 8, 16$ and $32$ GETT candidates across the benchmark; the total amount of viable candidates varies from TC to TC and is denoted by the numbers centered at each bar. We observe that GETT's performance model works to a point where empirical search (among multiple candidates) becomes almost obsolete in most cases. Quantitatively speaking, even in the extreme case where one limits the search space to a single candidate (i.e., eliminating search), GETT using single precision (double precision) on average still attains 90.7% (92.3%) of the performance of the fastest candidate. Actively searching through as little as four candidates increases the average performance to 98.3% (97.2%) for single precision (double precision). Moreover, we observe that searching through more than 16 candidates does not yield any performance benefit for any of the tested TCs.

---

[26]For instance, a real-valued GEMM of size $m = n = k$ has a theoretical arithmetic intensity of $\frac{2m^3}{3m^2} = \frac{2m}{3}$ while a GEMM of size $\tilde{m} = \tilde{n} = \tilde{k} = \frac{m}{2}$ only has an arithmetic intensity of $\frac{2\tilde{m}^3}{3\tilde{m}^2} = \frac{m}{3}$.

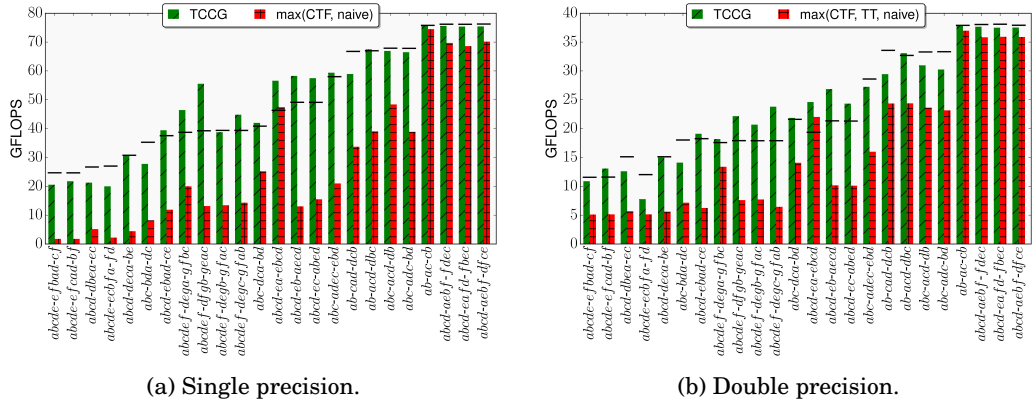

(a) Single precision.   (b) Double precision.

Fig. 12: TCCG performance. The horizontal black lines denote the performance of an equally-sized GEMM.

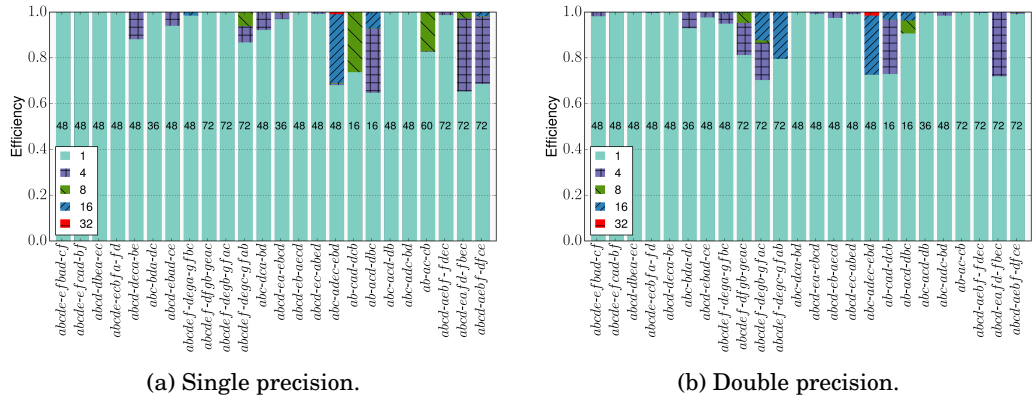

(a) Single precision.   (b) Double precision.

Fig. 13: Limit the number of GETT candidates to 1, 4, 8, 16, and 32. The numbers at the center of each bar represent the total amount of possible GETT candidates.

To give users the flexibility to trade GETT-performance for search time, TCCG offers the possibility to limit the search via the `--maxImplementations` command-line argument (default: 16 candidates).

## 8. CONCLUSION AND FUTURE WORK

We presented the GEMM-like Tensor-Tensor multiplication, a novel high-performance approach to compute tensor contractions. Several key features differentiate GETT from previous approaches:

—Any tensor contraction is reduced to a highly tuned and fully vectorized macro-kernel for which the operands of the macro-kernel are packed into a specified level of the cache hierarchy.
—The stride-one index is preserved throughout the packing process, resulting in a favourable memory access pattern.
—The arithmetic intensity with respect to an equally-sized GEMM is maintained.
—No additional auxiliary workspace is required.

We assessed GETT's performance on a benchmark for tensor contractions spanning a wide range of test cases; on average, GETT attains $98.1\%$ (minimum: $72.4\%$; maximum: $141.4\%$) and $97.0\%$ (minimum: $60.8\%$; maximum: $132.9\%$) of MKL's GEMM performance for single and double precision, respectively. In the compute-bound regime, GETT achieves an efficiency of up to $91.3\%$ of peak floating-point performance; on bandwidth-bound TCs, the positive effects of GETT's favorable memory-access patterns and its ability to block for the various cache levels are especially apparent, outperforming the existing TCs approaches by up to $12.4\times$.

To further assess GETT's performance, we carried out a thorough survey including two alternative approaches: Transpose-Transpose-GEMM-Transpose (TTGT) and Loops-over-GEMM (LoG). The survey exposes TTGT's shortcomings for bandwidth-bound TCs and LoG's arbitrarily poor performance in certain situations. While TTGT slightly outperforms GETT for compute-bound TCs (at the expense of additional memory), we do not anticipate any conceptual obstacles to eliminate this difference in a future implementation of GETT. In light of these results, we argue that a specialized, GEMM-like approach—such as GETT and TBLIS—has universal appeal, as it avoids the drawbacks of the previous methods—namely, additional memory requirement and suboptimal arithmetic intensity—and does not depend on the existence of a high-performance GEMM.

By combining GETT, TTGT and LoG into a unified Tensor Contraction Code Generator, we obtain high performance for bandwidth-bound and compute-bound tensor contractions alike. While TCCG offers the possibility to automatically search through multiple candidates, we found that the search space can be effectively pruned via architecture-aware metric (e.g., assuming knowledge about cache sizes). For instance, even in the extreme case where GETT's search is limited to a single candidate (that is, no search at all), GETT still attains $89.5\%$ and $87.7\%$ of the fastest implementation for single and double precision, respectively.

A multi-threaded version of GETT is available on GitHub. The parallelization follows closely the approach of the BLIS library [Smith et al. 2014], and is discussed in detail in the upcoming doctoral dissertation of the first author.

In the near future we would like to study the performance on Intel's latest Xeon Phi architecture (Knights Landing) and explore how our approach carries over to ternary tensor contractions of the form $\mathcal{D}_{\Pi^{\mathcal{D}}(I_m \cup I_n \cup I_l)} \leftarrow \alpha \times \mathcal{A}_{\Pi^{\mathcal{A}}(I_m \cup I_k)} \times \mathcal{B}_{\Pi^{\mathcal{B}}(I_n \cup I_k)} \times \mathcal{C}_{\Pi^{\mathcal{C}}(I_l \cup I_k)} + \beta \times \mathcal{D}_{\Pi^{\mathcal{D}}(I_m \cup I_n \cup I_l)}$, where contracted indices appear in three tensors simultaneously.

### ACKNOWLEDGMENT

Financial support from the Deutsche Forschungsgemeinschaft (DFG) through grant GSC 111 is gratefully acknowledged. Furthermore, we thank our colleagues in the HPAC research group for many fruitful discussions. We also thank our anonymous reviewers for helping us to improve the quality of this publication.

## A. FULL BENCHMARK

| Tensor contraction | Sizes | GETT | LoG | TTGT | CTF |
|---|---|---|---|---|---|
| *abc-bda-dc* | a:384;c:24;b:384;d:384; | 27.76 | - | 22.42 | 8.22 |
| *abc-dca-bd* | a:384;c:376;b:24;d:384; | 36.55 | 41.97 | 38.10 | 25.21 |
| *abcd-dbea-ec* | a:96;c:24;b:84;e:96;d:96; | 21.26 | - | 18.71 | 5.16 |
| *abcd-deca-be* | a:96;c:84;b:24;e:84;d:96; | 30.52 | - | 18.36 | 4.41 |
| *abcd-ebad-ce* | a:96;c:24;b:84;e:96;d:84; | 39.43 | 24.87 | 31.25 | 11.85 |
| *abcde-efbad-cf* | a:48;c:24;b:36;e:48;d:36;f:36; | 20.51 | - | 14.85 | 1.75 |
| *abcde-ecbfa-fd* | a:48;c:36;b:36;e:48;d:24;f:48; | 19.99 | - | 15.77 | 2.22 |
| *abcde-efcad-bf* | a:48;c:36;b:24;e:48;d:36;f:36; | 21.76 | - | 13.82 | 1.75 |
| *abcd-ea-ebcd* | a:96;c:84;b:84;e:96;d:84; | 56.58 | 46.30 | 46.31 | 47.40 |
| *abcd-eb-aecd* | a:96;c:84;b:84;e:96;d:84; | 58.26 | 45.05 | 35.43 | 13.00 |
| *abcd-ec-abed* | a:96;c:84;b:84;e:96;d:84; | 57.46 | 49.45 | 35.36 | 15.46 |
| *ab-ac-cb* | a:7248;c:7248;b:7240; | 65.89 | 75.80 | 75.80 | 74.45 |
| *ab-acd-dbc* | a:384;c:376;b:376;d:384; | 65.12 | 67.46 | 60.02 | 39.05 |
| *ab-cad-dcb* | a:384;c:384;b:376;d:384; | 48.31 | - | 58.90 | 33.68 |
| *abc-acd-db* | a:384;c:376;b:376;d:384; | 66.90 | 63.17 | 63.87 | 48.35 |
| *abc-ad-bdc* | a:384;c:376;b:384;d:376; | 68.92 | 63.60 | 60.57 | 48.61 |
| *abc-adc-bd* | a:384;c:376;b:384;d:376; | 66.49 | 63.98 | 60.25 | 38.87 |
| *abc-adc-db* | a:384;c:376;b:376;d:384; | 66.81 | 63.65 | 60.19 | 38.86 |
| *abc-adec-ebd* | a:96;c:84;b:84;e:96;d:84; | 59.34 | 50.65 | 45.37 | 20.98 |
| *abcd-aebf-dfce* | a:96;c:84;b:84;e:84;d:96;f:84; | 67.85 | 57.83 | 75.44 | 70.13 |
| *abcd-aebf-fdec* | a:96;c:84;b:84;e:84;d:84;f:96; | 67.57 | 57.68 | 75.57 | 69.67 |
| *abcd-aecf-bfde* | a:96;c:84;b:96;e:84;d:84;f:84; | 70.15 | 59.11 | 75.42 | 70.00 |
| *abcd-aecf-fbed* | a:96;c:84;b:84;e:84;d:84;f:96; | 69.62 | 58.46 | 75.40 | 69.52 |
| *abcd-aedf-bfce* | a:96;c:84;b:96;e:84;d:84;f:84; | 70.56 | 59.03 | 75.39 | 69.83 |
| *abcd-aedf-fbec* | a:96;c:84;b:84;e:84;d:84;f:96; | 69.78 | 58.51 | 75.56 | 69.32 |
| *abcd-aefb-fdce* | a:96;c:84;b:84;e:84;d:84;f:96; | 67.51 | 57.69 | 75.80 | 69.45 |
| *abcd-aefc-fbed* | a:96;c:84;b:84;e:84;d:84;f:96; | 69.45 | 58.66 | 75.54 | 69.31 |
| *abcd-eafb-fdec* | a:96;c:84;b:84;e:96;d:84;f:96; | 69.08 | - | 75.64 | 68.76 |
| *abcd-eafc-bfde* | a:96;c:84;b:96;e:96;d:84;f:84; | 69.85 | 58.72 | 75.67 | 69.05 |
| *abcd-eafd-fbec* | a:96;c:84;b:84;e:96;d:84;f:96; | 70.62 | - | 75.36 | 68.58 |
| *abcdef-dega-gfbc* | a:24;c:20;b:20;e:20;d:24;g:24;f:20; | 46.41 | - | 22.55 | 3.16 |
| *abcdef-degb-gfac* | a:24;c:20;b:20;e:20;d:24;g:24;f:20; | 38.74 | 20.71 | 24.02 | 2.89 |
| *abcdef-degc-gfab* | a:24;c:20;b:20;e:20;d:24;g:24;f:20; | 44.83 | 38.49 | 25.23 | 2.87 |
| *abcdef-dfga-gebc* | a:24;c:20;b:20;e:20;d:24;g:24;f:20; | 46.74 | - | 23.39 | 3.01 |
| *abcdef-dfgb-geac* | a:24;c:20;b:20;e:20;d:24;g:24;f:20; | 55.52 | 26.35 | 24.83 | 2.75 |
| *abcdef-dfgc-geab* | a:24;c:20;b:20;e:20;d:24;g:24;f:20; | 43.33 | 18.83 | 25.24 | 2.74 |
| *abcdef-efga-gdbc* | a:24;c:20;b:20;e:24;d:20;g:24;f:20; | 46.75 | - | 23.25 | 2.43 |
| *abcdef-efgb-gdac* | a:24;c:20;b:20;e:24;d:20;g:24;f:20; | 35.73 | 19.37 | 25.01 | 2.12 |
| *abcdef-efgc-gdab* | a:24;c:20;b:20;e:24;d:20;g:24;f:20; | 47.08 | 38.30 | 25.37 | 2.12 |
| *abcdef-gdab-efgc* | a:24;c:20;b:20;e:24;d:20;g:24;f:20; | 47.48 | 34.59 | 25.22 | 2.12 |
| *abcdef-gdac-efgb* | a:24;c:20;b:20;e:24;d:20;g:24;f:20; | 35.74 | 19.57 | 24.25 | 2.11 |
| *abcdef-gdbc-efga* | a:24;c:20;b:20;e:24;d:20;g:24;f:20; | 46.80 | - | 23.34 | 2.42 |
| *abcdef-geab-dfgc* | a:24;c:20;b:20;e:20;d:24;g:24;f:20; | 48.26 | 19.53 | 24.88 | 2.74 |
| *abcdef-geac-dfgb* | a:24;c:20;b:20;e:20;d:24;g:24;f:20; | 55.69 | 28.12 | 23.34 | 2.76 |
| *abcdef-gebc-dfga* | a:24;c:20;b:20;e:20;d:24;g:24;f:20; | 46.76 | - | 22.86 | 3.01 |
| *abcdef-gfab-degc* | a:24;c:20;b:20;e:20;d:24;g:24;f:20; | 45.34 | 34.85 | 24.67 | 2.87 |
| *abcdef-gfac-degb* | a:24;c:20;b:20;e:20;d:24;g:24;f:20; | 38.78 | 21.03 | 22.99 | 2.89 |
| *abcdef-gfbc-dega* | a:24;c:20;b:20;e:20;d:24;g:24;f:20; | 46.43 | - | 24.45 | 3.16 |

Table III: Full benchmark results (in GFLOPS) using single precision.

| Tensor contraction | Sizes | GETT | LoG | TTGT | CTF |
|---|---|---|---|---|---|
| $abc\text{-}bda\text{-}dc$ | a:312;c:24;b:312;d:312; | 14.10 | - | 11.27 | 4.44 |
| $abc\text{-}dca\text{-}bd$ | a:312;c:296;b:24;d:312; | 17.58 | 21.86 | 20.01 | 14.07 |
| $abcd\text{-}dbea\text{-}ec$ | a:72;c:24;b:72;e:72;d:72; | 12.60 | - | 9.14 | 2.97 |
| $abcd\text{-}deca\text{-}be$ | a:72;c:72;b:24;e:72;d:72; | 14.96 | - | 9.24 | 2.86 |
| $abcd\text{-}ebad\text{-}ce$ | a:72;c:24;b:72;e:72;d:72; | 19.11 | 12.61 | 15.19 | 6.23 |
| $abcde\text{-}efbad\text{-}cf$ | a:48;c:24;b:32;e:48;d:32;f:32; | 10.87 | - | 7.05 | 1.25 |
| $abcde\text{-}ecbfa\text{-}fd$ | a:48;c:32;b:32;e:48;d:24;f:48; | 7.31 | - | 7.77 | 1.74 |
| $abcde\text{-}efcad\text{-}bf$ | a:48;c:32;b:24;e:48;d:32;f:32; | 13.08 | - | 6.75 | 1.25 |
| $abcd\text{-}ea\text{-}ebcd$ | a:72;c:72;b:72;e:72;d:72; | 24.59 | 19.36 | 19.36 | 22.02 |
| $abcd\text{-}eb\text{-}aecd$ | a:72;c:72;b:72;e:72;d:72; | 26.86 | 19.75 | 15.00 | 6.12 |
| $abcd\text{-}ec\text{-}abed$ | a:72;c:72;b:72;e:72;d:72; | 24.31 | 23.04 | 15.05 | 6.46 |
| $ab\text{-}ac\text{-}cb$ | a:5136;c:5136;b:5120; | 31.37 | 37.89 | 37.89 | 36.95 |
| $ab\text{-}acd\text{-}dbc$ | a:312;c:296;b:296;d:312; | 28.44 | 33.06 | 28.80 | 20.03 |
| $ab\text{-}cad\text{-}dcb$ | a:312;c:312;b:296;d:312; | 28.41 | - | 29.42 | 17.70 |
| $abc\text{-}acd\text{-}db$ | a:312;c:296;b:296;d:312; | 29.29 | 30.09 | 30.95 | 23.18 |
| $abc\text{-}ad\text{-}bdc$ | a:312;c:296;b:312;d:296; | 30.33 | 29.95 | 28.55 | 22.97 |
| $abc\text{-}adc\text{-}bd$ | a:312;c:296;b:312;d:296; | 29.39 | 30.24 | 28.81 | 18.01 |
| $abc\text{-}adc\text{-}db$ | a:312;c:296;b:296;d:312; | 29.24 | 30.28 | 28.79 | 17.99 |
| $abc\text{-}adec\text{-}ebd$ | a:72;c:72;b:72;e:72;d:72; | 27.24 | 25.30 | 21.78 | 10.66 |
| $abcd\text{-}aebf\text{-}dfce$ | a:72;c:72;b:72;e:72;d:72;f:72; | 30.93 | 25.38 | 37.50 | 34.15 |
| $abcd\text{-}aebf\text{-}fdec$ | a:72;c:72;b:72;e:72;d:72;f:72; | 30.77 | 26.26 | 37.65 | 31.81 |
| $abcd\text{-}aecf\text{-}bfde$ | a:72;c:72;b:72;e:72;d:72;f:72; | 32.02 | 27.09 | 37.34 | 34.09 |
| $abcd\text{-}aecf\text{-}fbed$ | a:72;c:72;b:72;e:72;d:72;f:72; | 31.64 | 28.00 | 37.46 | 32.80 |
| $abcd\text{-}aedf\text{-}bfce$ | a:72;c:72;b:72;e:72;d:72;f:72; | 32.08 | 27.16 | 37.33 | 34.07 |
| $abcd\text{-}aedf\text{-}fbec$ | a:72;c:72;b:72;e:72;d:72;f:72; | 31.65 | 27.99 | 37.46 | 32.99 |
| $abcd\text{-}aefb\text{-}fdce$ | a:72;c:72;b:72;e:72;d:72;f:72; | 30.77 | 26.66 | 37.64 | 31.82 |
| $abcd\text{-}aefc\text{-}fbed$ | a:72;c:72;b:72;e:72;d:72;f:72; | 31.53 | 28.38 | 37.45 | 32.76 |
| $abcd\text{-}eafb\text{-}fdec$ | a:72;c:72;b:72;e:72;d:72;f:72; | 30.59 | - | 37.69 | 33.71 |
| $abcd\text{-}eafc\text{-}bfde$ | a:72;c:72;b:72;e:72;d:72;f:72; | 31.89 | 26.92 | 37.37 | 33.62 |
| $abcd\text{-}eafd\text{-}fbec$ | a:72;c:72;b:72;e:72;d:72;f:72; | 31.63 | - | 37.48 | 33.68 |
| $abcdef\text{-}dega\text{-}gfbc$ | a:24;c:16;b:16;e:16;d:24;g:24;f:16; | 18.19 | - | 11.53 | 1.43 |
| $abcdef\text{-}degb\text{-}gfac$ | a:24;c:16;b:16;e:16;d:24;g:24;f:16; | 20.69 | 8.81 | 11.59 | 1.21 |
| $abcdef\text{-}degc\text{-}gfab$ | a:24;c:16;b:16;e:16;d:24;g:24;f:16; | 23.80 | 18.13 | 11.84 | 1.21 |
| $abcdef\text{-}dfga\text{-}gebc$ | a:24;c:16;b:16;e:16;d:24;g:24;f:16; | 18.19 | - | 11.49 | 1.42 |
| $abcdef\text{-}dfgb\text{-}geac$ | a:24;c:16;b:16;e:16;d:24;g:24;f:16; | 22.15 | 7.79 | 11.79 | 1.19 |
| $abcdef\text{-}dfgc\text{-}geab$ | a:24;c:16;b:16;e:16;d:24;g:24;f:16; | 24.00 | 11.75 | 11.85 | 1.20 |
| $abcdef\text{-}efga\text{-}gdbc$ | a:24;c:16;b:16;e:24;d:16;g:24;f:16; | 18.29 | - | 11.17 | 1.25 |
| $abcdef\text{-}efgb\text{-}gdac$ | a:24;c:16;b:16;e:24;d:16;g:24;f:16; | 18.88 | 7.83 | 11.84 | 1.18 |
| $abcdef\text{-}efgc\text{-}gdab$ | a:24;c:16;b:16;e:24;d:16;g:24;f:16; | 24.12 | 16.10 | 11.86 | 1.19 |
| $abcdef\text{-}gdab\text{-}efgc$ | a:24;c:16;b:16;e:24;d:16;g:24;f:16; | 23.97 | 16.04 | 11.82 | 1.19 |
| $abcdef\text{-}gdac\text{-}efgb$ | a:24;c:16;b:16;e:24;d:16;g:24;f:16; | 18.97 | 8.00 | 11.64 | 1.19 |
| $abcdef\text{-}gdbc\text{-}efga$ | a:24;c:16;b:16;e:24;d:16;g:24;f:16; | 18.31 | - | 10.97 | 1.24 |
| $abcdef\text{-}geab\text{-}dfgc$ | a:24;c:16;b:16;e:16;d:24;g:24;f:16; | 24.26 | 13.00 | 11.79 | 1.20 |
| $abcdef\text{-}geac\text{-}dfgb$ | a:24;c:16;b:16;e:16;d:24;g:24;f:16; | 22.27 | 7.95 | 11.54 | 1.20 |
| $abcdef\text{-}gebc\text{-}dfga$ | a:24;c:16;b:16;e:16;d:24;g:24;f:16; | 18.25 | - | 11.34 | 1.42 |
| $abcdef\text{-}gfab\text{-}degc$ | a:24;c:16;b:16;e:16;d:24;g:24;f:16; | 24.62 | 16.94 | 11.80 | 1.21 |
| $abcdef\text{-}gfac\text{-}degb$ | a:24;c:16;b:16;e:16;d:24;g:24;f:16; | 20.69 | 8.93 | 11.30 | 1.21 |
| $abcdef\text{-}gfbc\text{-}dega$ | a:24;c:16;b:16;e:16;d:24;g:24;f:16; | 18.19 | - | 11.35 | 1.44 |

Table IV: Full benchmark results (in GFLOPS) using double precision.